\documentclass[usenatbib]{mn2e}
\usepackage{graphicx,amsmath,multirow,amssymb}
\usepackage{subfig}
\usepackage{natbib}
\newcommand{\comment}[1]{}

\def\simgt{\lower.5ex\hbox{$\; \buildrel > \over \sim \;$}}
\def\simlt{\lower.5ex\hbox{$\; \buildrel < \over \sim \;$}}

\newcommand{\Msun}{\ensuremath{{\rm M}_{\sun}}}

\title[Solar metallicity AGBs]{Studying the evolution of AGB stars in the Gaia epoch}
\author[Di Criscienzo al.]{M. Di Criscienzo$^1$, P. Ventura$^1$,  
D. A. Garc\'{\i}a--Hern\'andez$^{2,3}$, F. Dell'Agli$^{1}$,
\newauthor
M. Castellani$^{1}$, P. M. Marrese$^{1,4}$, S. Marinoni$^{1,4}$, G. Giuffrida$^{1,4}$, O. Zamora$^{2,3}$   \\ \\
$^1$INAF -- Osservatorio Astronomico di Roma, Via Frascati 33, 00040, Monte Porzio Catone (RM), Italy \\
$^{2}$Instituto de Astrof\'{\i}sica de Canarias, E-38205 La Laguna, Tenerife, Spain \\
$^{3}$Departamento de Astrof\'{\i}sica, Universidad de La Laguna (ULL), E-38206 La Laguna, Tenerife, Spain\\
$^{4}$ASDC-ASI, Via del Politecnico, I-00133 Roma, Italy
}

\begin{document}

\date{Accepted, Received; in original form }

\pagerange{\pageref{firstpage}--\pageref{lastpage}} \pubyear{2012}

\maketitle

\label{firstpage}

\begin{abstract}
We present asymptotic giant branch (AGB) models of solar metallicity, to allow the 
interpretation of observations of Galactic AGB stars, whose distances should 
be soon available after the first release of the Gaia catalogue.
We find an abrupt change in the AGB physical and chemical properties, occurring at the threshold mass 
 to ignite hot bottom burning,i.e. $3.5~\Msun$. Stars
with mass below $3.5~\Msun$ reach the C-star stage and eject into the interstellar medium gas enriched in carbon , nitrogen and $^{17}O$. 
The higher mass counterparts evolve at large luminosities, between $3\times 10^4 L_{\odot}$ and $10^5 L_{\odot}$. 
The mass expelled from the massive AGB stars
shows the imprinting of proton-capture nucleosynthesis, with considerable 
production of nitrogen and sodium and destruction of $^{12}C$ and $^{18}O$. 
The comparison with the most recent results from other research groups are discussed, to evaluate the
robustness of the present findings. Finally, we compare the models 
with recent observations
of galactic AGB stars, outlining the possibility offered by Gaia to shed new light on
the evolution properties of this class of objects.
\end{abstract}


\section{Introduction}
Stars of mass in the range $1\Msun \leq M \leq 8\Msun$, after the consumption of helium in the core,
evolve through the asymptotic giant branch phase: above a degenerate core,
composed of carbon and oxygen (or of oxygen and neon, in the stars of highest mass),
a $3\alpha$ burning zone and a region with CNO nuclear activity provide alternatively
the energy required to support the star \citep{becker80, iben82, iben83, lattanzio86}.
Because helium burning is activated in
condition of thermal instability \citep{schw65, schw67}, CNO cycling is for most of the 
time the only active nuclear channel, whereas ignition of helium occurs periodically, during rapid events, known as thermal pulses (TP).\\

Though the duration of the AGB phase is extremely short when compared to the evolutionary
time of the star, it proves of paramount importance for the feedback of these stars
on the host environment. This is because it is during the AGB evolution that intermediate
mass stars lose their external mantle, thus contributing to the gas pollution of the
interstellar medium. In addition, these stars have been recognised as important 
manufacturers of dust, owing to the thermodynamic conditions of their winds, which are a 
favourable environment to the condensation of gas molecules into solid particles 
\citep{gail99}.\\
For the above reasons, AGB stars are believed to play a crucial role in several 
astrophysical contexts. \\
On a pure stellar evolution side, they are an ideal laboratory to 
test stellar evolution theories, because of the complexity of their internal structure.
In the context of the Galaxy evolution, the importance of AGB stars for the determination
of the chemical trends traced by stars in different parts of the Milky Way has been
recognised in several studies \citep{romano10, kobayashi11}. Still in the Milky Way
environment, massive AGB stars have been proposed as the main actors in the formation of 
multiple populations in Globular Clusters \citep{ventura01}. Moving out to the Galaxy,
it is generally believed that AGB stars give an important contribution to the dust present 
at high redshift \citep{valiante09, valiante11}; furthermore, these stars play a 
crucial role in the formation and evolution of galaxies (Santini et al. 2014).\\
It is for these reasons that the research on AGB stars has attracted the interests of the
astrophysical community in the last decades.\\
The description of these stars is extremely difficult, owing to the very short time steps (of the order of one day)
required to describe the TP phases, which leads to very long computation times. 
Furthermore, the evolutionary properties of these stars are determined by the delicate 
interface between the degenerate core and the tenuous, expanded envelope, thus rendering 
the results obtained extremely sensitive to convection modelling \citep{herwig05, karakas14b}. \\
There are two mechanisms potentially able to alter the surface chemical 
composition, namely hot bottom-burning (hereinafter HBB) and third dregde-up (TDU).
The efficiency of the two mechanisms potentially able to alter the surface chemical 
composition, namely hot bottom burning (hereinafter HBB) and third dregde-up (TDU) 
depends critically on the method used to determine the temperature gradients in regions 
unstable to convective motions \citep{vd05a} and on the details of the treatment of 
the convective borders, for what concerns
the base of the convective envelope and the boundaries of the shell that forms 
in conjunction with each TP, the so called "pulse driven convective shell".
The description of mass loss also plays an important role in the determination of the
evolutionary time scales \citep{vd05b, doherty14}.\\
Given the poor knowledge of some of the macro-physics input necessary to build the
evolutionary sequences, primarily convection and mass loss, the comparison with the
observations is at the moment the only way to improve the robustness of the results
obtained. \\
On this side, the Magellanic Clouds have been so far used much more extensively
than the Milky Way \citep{martin93, marigo99, karakas02, izzard04, marigo07, stancliffe05}, 
given the unknown distances of Galactic sources, which render difficult any interpretation 
of the observations. Very recent works outlines the possibility of calibrating AGB models 
based on the observations of the AGB population in dwarf galaxies in the Local Group
\citep{rosenfield14, rosenfield16}.
The attempts of interpreting the observations of metal poor 
environments, typical of the Magellanic Clouds and of the galaxies in the Local Group,
has so far pushed our attention towards sub-solar AGB models, published in previous works
of our group \citep{ventura08, ventura09, ventura11, ventura13}. The main drivers of
these researches were the understanding of the presence of multiple populations in
globular clusters 
%(see for example Ventura et al. 2012,2014)
and the comparison of our predictions with the evolved stellar
population of the Magellanic Clouds \citep{flavia15a, flavia15b, ventura15, ventura16} and metal poor dwarf galaxies of Local Group \citep{dellagli16}.
%%%% GAIA
The advent of the ESA-Gaia mission will open new frontiers in the study of stars of any 
class, and in particular for the evolved stellar population of the Milky Way. Launched 
on December 2013, Gaia will allow constructing a catalogue of around more than 1 billion astronomical 
objects (mostly stars) brighter than 20 G mag (where G is the Gaia whitelight passband, Jordi et al. 2010), 
which encompasses $\sim 1\%$ of the Galactic stellar population. During the five year 
mission life time each  object will be observed  70 times on average, for a total  of 
$\sim 630$  photometric measurements in G band, the exact number of observations depending 
on the magnitude and position of the object (ecliptic coordinates) and on the stellar 
density in the object field. Gaia will perform $\mu$as global astrometry for all the 
observed objects, thus
allowing the determination of the distance of several AGB stars with unprecedented accuracy,
refining the parallaxes determination of all the stars in the 
Hipparcos catalogue and dramatically increasing the number of accurately known
parallaxes. The first release of the Gaia catalogue is foreseen  by the end of summer 2016, 
and it will contain  positions and G-magnitudes for all single objects with good
astrometric behaviour. 
%%%%
In order to benefit from the possibilities offered by the upcoming Gaia data, we calculated new AGB models with 
solar metallicity, completing our library, so far 
limited to sub-solar chemical composition models. The main goal of the present work is to 
explore the possibilities, offered by the comparisons with observations, to further 
constrain some of the still poorly known phenomena affecting this class of objects. 
This task is essential to be able to assess the role played by AGB stars in the various 
contexts discussed earlier in the section.\\
To this aim, after the presentation of the main physical and chemical properties of the solar chemistry AGB models, 
we will compare our theoretical results with  a) the models available in the literature, to determine their degree of uncertainty 
and their robustness and  b) recent observations of galactic AGB. In some cases we will also discuss how Gaia will help 
discriminating among various possibilities still open at present.\\

The paper is structured as follows: the description of the input used to build the 
evolutionary sequences is given in section 2; in section 3 we present an overall review of
the evolution through the AGB 
ADD phase; 
the contamination of the interstellar medium determined by the gas  ejected 
from these stars is discussed in section 4; section 5 presents a
detailed comparison with two among the most largely used sets of models available in the
literature; in section 6 we test our models against the chemical composition of samples
of Galactic AGB stars; the conclusions are given in section 7.

\section{Physical and chemical input}

The evolutionary sequences used in this work were calculated with the ATON code; the
details of the numerical and physical characteristics of the code are thoroughly documented
in \citet{ventura98}, while the most recent updates are presented in \citet{ventura09}.
The interested reader is addressed to those papers for the details of the
input adopted to build the evolutionary sequences. Here we provide the ingredients 
most relevant for the present analysis:
\begin{itemize}
\item{
{\it Chemical composition.} The models presented here are representative of the solar
chemical composition. The metallicity is $Z=0.017$, with initial helium $Y=0.28$. The distribution 
of the different chemical elements in the initial mixture is taken from \citet{gs98}. 
}
\item{
{\it Mass range}. The initial mass values are between $1\Msun$ and  $8\Msun$. 
We did not consider initial masses below $1.25\Msun$, as their surface chemical composition is contaminated only by
the first dredge-up, with scarce modification from TDU and no effects from HBB;
the chemistry of the  $\le$ $1.25\Msun$ model  reflect a modest contribution from
TDU and never reaches the carbon star stage. On the other hand stars with initial mass above $8\Msun$ 
undergo core collapse, thus skipping the AGB phase.
}
\item{
{\it Convection.} 
In regions unstable to convective motions, the temperature gradient is determined via
the full spectrum of turbulence (FST) model \citep{cm91}. In convective zones where 
nuclear reactions are active we couple mixing of chemicals and nuclear burning in
a diffusive-like scheme \citep{cloutmann}. The overshoot from the
convective borders (fixed by the Schwarzschild criterion) is described by an exponential 
decay of convective velocities; the extent of the overshoot region is determined by the 
e-folding distance of such a decay, which in pressure scale height ($H_p$) units, is 
given by $\zeta \times H_p$. During the core hydrogen-burning phase of stars of mass 
$M\geq 1.5\Msun$, we assume an extra-mixing from the external border of the convective core, 
with $\zeta=0.02$; this is based on the constraint on core-overshoot necessary to reproduce the 
observed width of the main sequences of open clusters, given in \citet{ventura98}. The 
same overshoot is applied during the core helium burning phases of the stars of any mass.

During the AGB phase, we allow extra mixing from the internal border of the envelope and
from the boundaries of the pulse driven convective shell; we use 
$\zeta=0.002$, in agreement with the calibration based on the observed luminosity
function of carbon stars in the LMC, given by \citet{ventura14a}. 
}

\item{
{\it Mass loss.} The mass loss rate for oxygen-rich models is determined via the 
\citet{blocker95} treatment; the parameter entering the \citet{blocker95}'s recipe was
set to $\eta=0.02$, following \citet{ventura00}. Once the stars reach the C-star stage,
we use the description of mass loss from the Berlin group \citep{wachter02, wachter08}.
} 

\item{
{\it Opacities.} Radiative opacities are calculated according to the OPAL
release, in the version documented by \citet{opal}. The molecular opacities in the 
low-temperature regime ($T < 10^4$ K) are calculated by means of the AESOPUS tool \citep{marigo09}. 
The opacities are constructed to follow the changes of the envelop chemical composition,
in particular carbon, nitrogen, and oxygen individual abundances.
} 

\end{itemize}

\begin{table*}
\caption{AGB evolution properties of solar metallicity models}                                       
\begin{tabular}{c c c c c c c c}        
\hline\hline                        
$M/ \Msun$  &  $\tau_{AGB}$  &  $\%(C_{star})^a$  & $L_{max}/L_{\odot}$  &  $T_{bce}^{max}$ & 
$\lambda_{max}$$^b$  &  $M_C/\Msun$  &  $M_f/\Msun$  \\
\hline       
%1.0  & $3.1\times 10^6$ & 0  & $5.8\times 10^3$ & $2.8\times 10^6$ & 0.21 & 0.50 & 0.58  \\
1.25 & $2.0\times 10^6$ & 0  & $7.6\times 10^3$ & $3.4\times 10^6$ & 0.28 & 0.51 & 0.59  \\
1.5  & $2.1\times 10^6$ & 4  & $8.7\times 10^3$ & $4.3\times 10^6$ & 0.32 & 0.51 & 0.61  \\
1.75 & $2.4\times 10^6$ & 4  & $9.7\times 10^3$ & $5.6\times 10^6$ & 0.39 & 0.51 & 0.615 \\
2.0  & $3.8\times 10^6$ & 4  & $9.1\times 10^3$ & $6.0\times 10^6$ & 0.46 & 0.49 & 0.62  \\
2.25 & $4.2\times 10^6$ & 5  & $1.1\times 10^4$ & $7.8\times 10^6$ & 0.48 & 0.49 & 0.63  \\
2.5  & $4.0\times 10^6$ & 10 & $1.2\times 10^4$ & $8.0\times 10^6$ & 0.62 & 0.49 & 0.65  \\
3.0  & $1.8\times 10^6$ & 15 & $1.3\times 10^4$ & $1.1\times 10^7$ & 0.81 & 0.56 & 0.67  \\
3.5  & $1.0\times 10^6$ & 0  & $2.6\times 10^4$ & $7.1\times 10^7$ & 0.58 & 0.66 & 0.78  \\
4.0  & $3.1\times 10^5$ & 0  & $3.1\times 10^4$ & $8.0\times 10^7$ & 0.32 & 0.79 & 0.86  \\
4.5  & $2.4\times 10^5$ & 0  & $3.8\times 10^4$ & $8.4\times 10^7$ & 0.27 & 0.83 & 0.89  \\
5.0  & $1.9\times 10^5$ & 0  & $4.6\times 10^4$ & $8.7\times 10^7$ & 0.23 & 0.86 & 0.91  \\
5.5  & $1.6\times 10^5$ & 0  & $5.4\times 10^4$ & $8.9\times 10^7$ & 0.21 & 0.89 & 0.94  \\
6.0  & $1.2\times 10^5$ & 0  & $6.3\times 10^4$ & $9.1\times 10^7$ & 0.18 & 0.93 & 0.97  \\
6.5  & $9.0\times 10^4$ & 0  & $7.4\times 10^4$ & $9.4\times 10^7$ & 0.13 & 0.99 & 1.02  \\
7.0  & $8.0\times 10^4$ & 0  & $8.7\times 10^4$ & $1.0\times 10^8$ & 0    & 1.04 & 1.08  \\
7.5  & $6.0\times 10^4$ & 0  & $9.5\times 10^4$ & $1.0\times 10^8$ & 0    & 1.14 & 1.16  \\
8.0  & $5.0\times 10^4$ & 0  & $1.0\times 10^5$ & $1.1\times 10^8$ & 0    & 1.21 & 1.25  \\
\hline     
\end{tabular}

$^a$ Percentage of the duration of the C-rich phase;$^b$ the maximum efficiency of TDU, defined as the ratio between the mass 
mixed  in
the surface convection region and the mass processed by CNO burning during the interpulse phase

\end{table*}

\section{The evolution through the AGB phase}

\begin{figure*}
\begin{minipage}{0.48\textwidth}
%\resizebox{1.\hsize}{!}{\includegraphics{figlum.ps}}
\resizebox{1.\hsize}{!}{\includegraphics{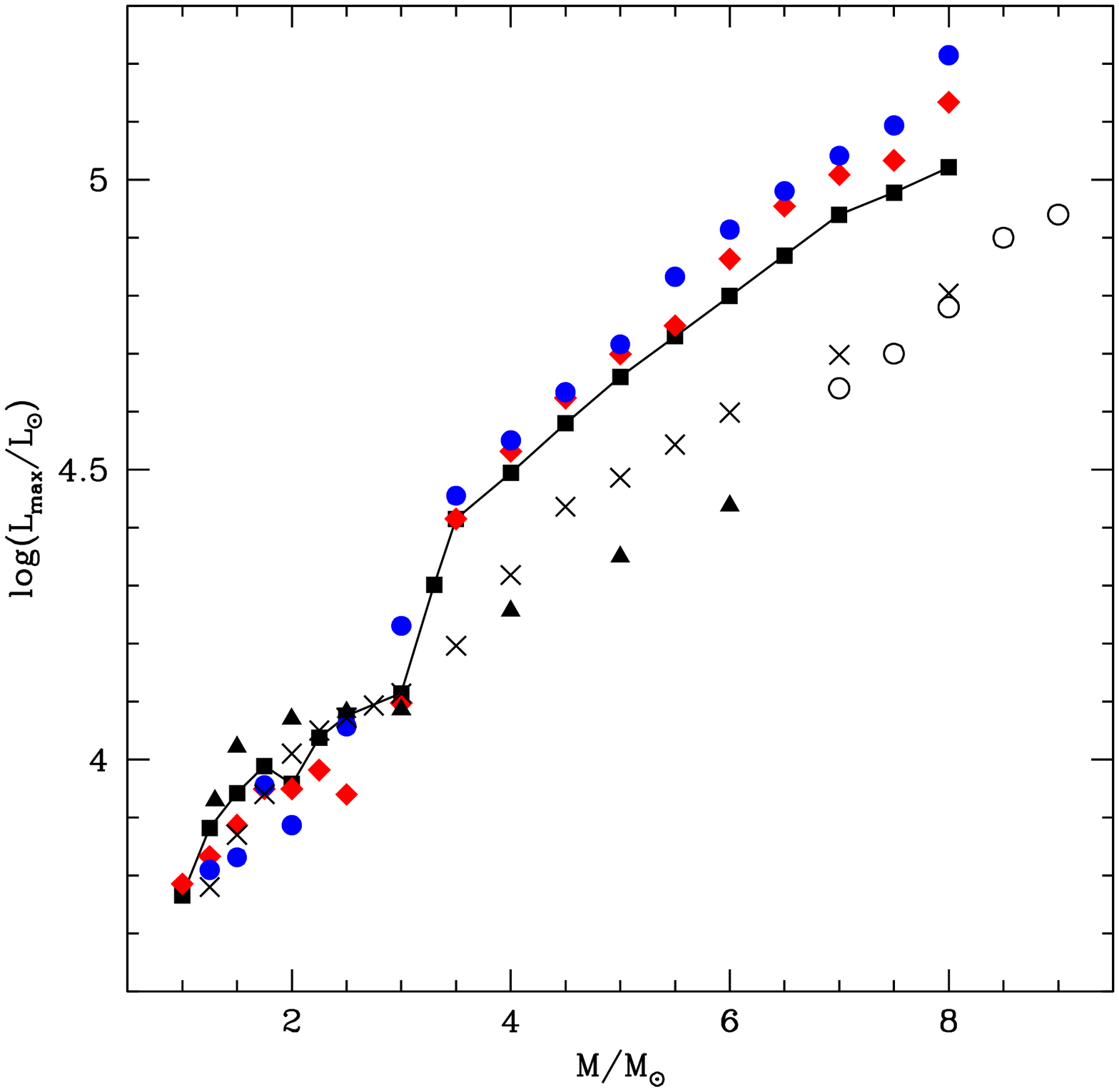}}
\end{minipage}
\begin{minipage}{0.48\textwidth}
%\resizebox{1.\hsize}{!}{\includegraphics{figtbce.ps}}
\resizebox{1.\hsize}{!}{\includegraphics{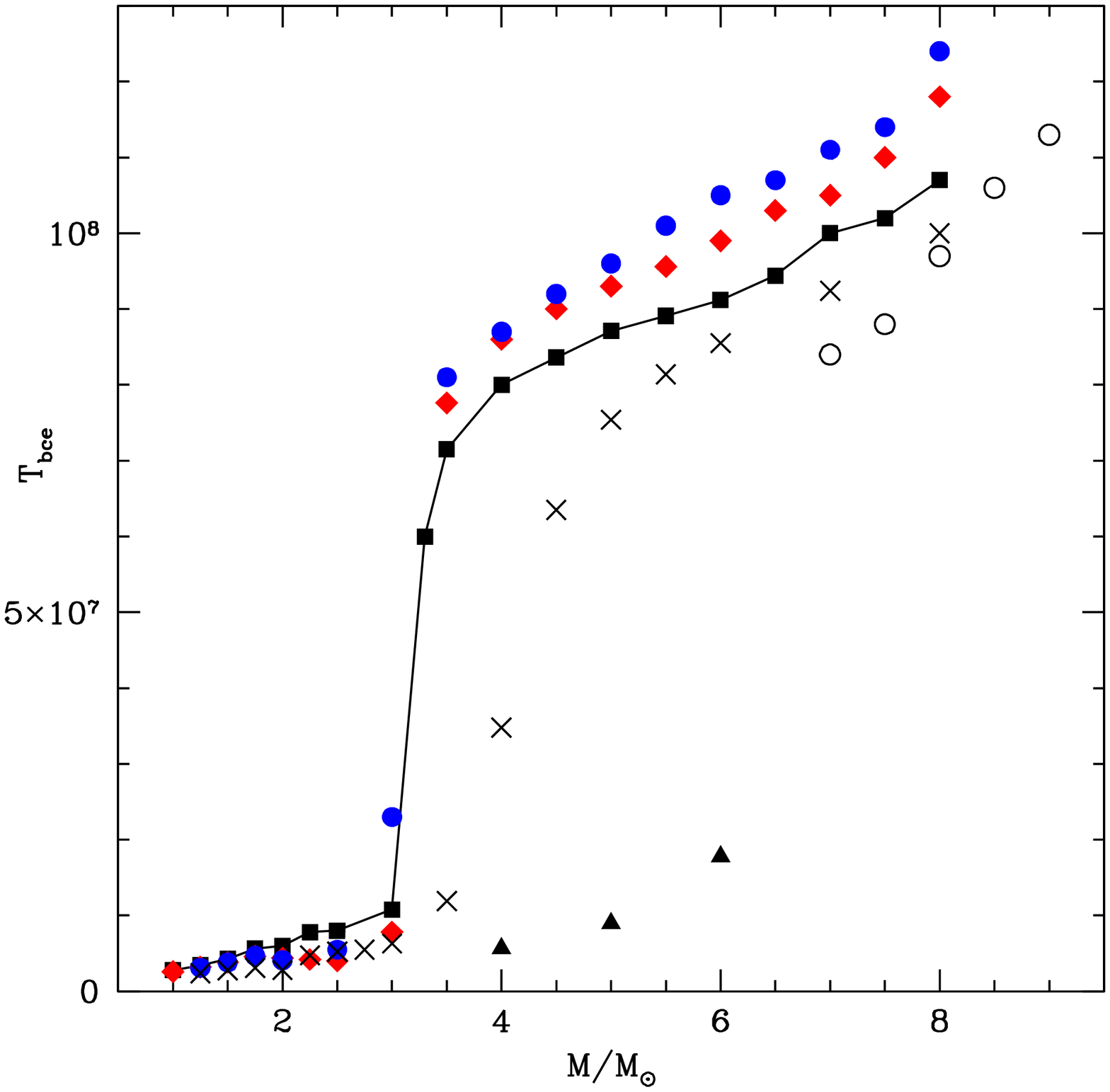}}
\end{minipage}
%\vskip-80pt
\begin{minipage}{0.48\textwidth}
%\resizebox{1.\hsize}{!}{\includegraphics{figmcore.ps}}
\resizebox{1.\hsize}{!}{\includegraphics{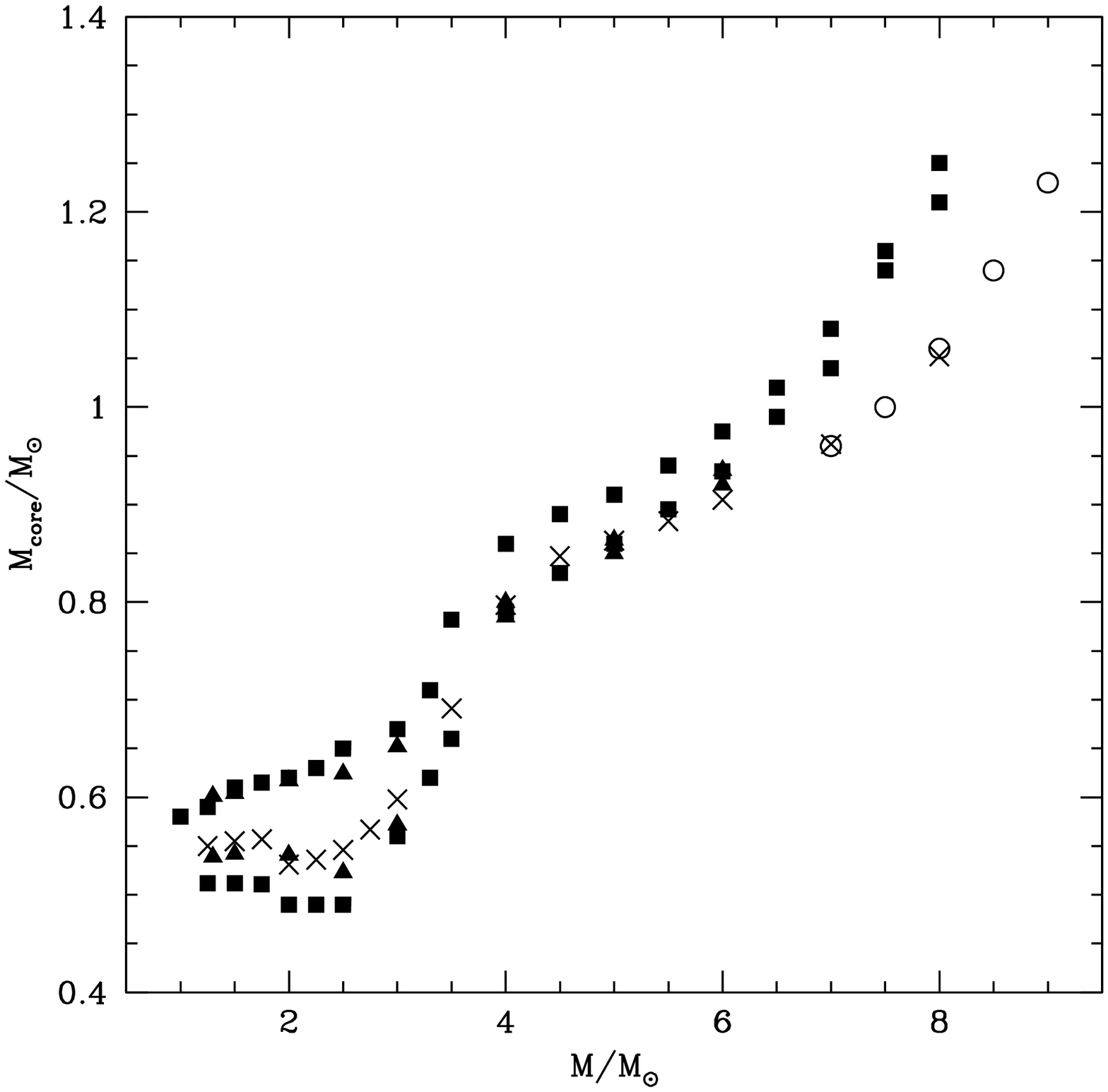}}
\end{minipage}
\begin{minipage}{0.48\textwidth}
%\resizebox{1.\hsize}{!}{\includegraphics{figtau.ps}}
\resizebox{1.\hsize}{!}{\includegraphics{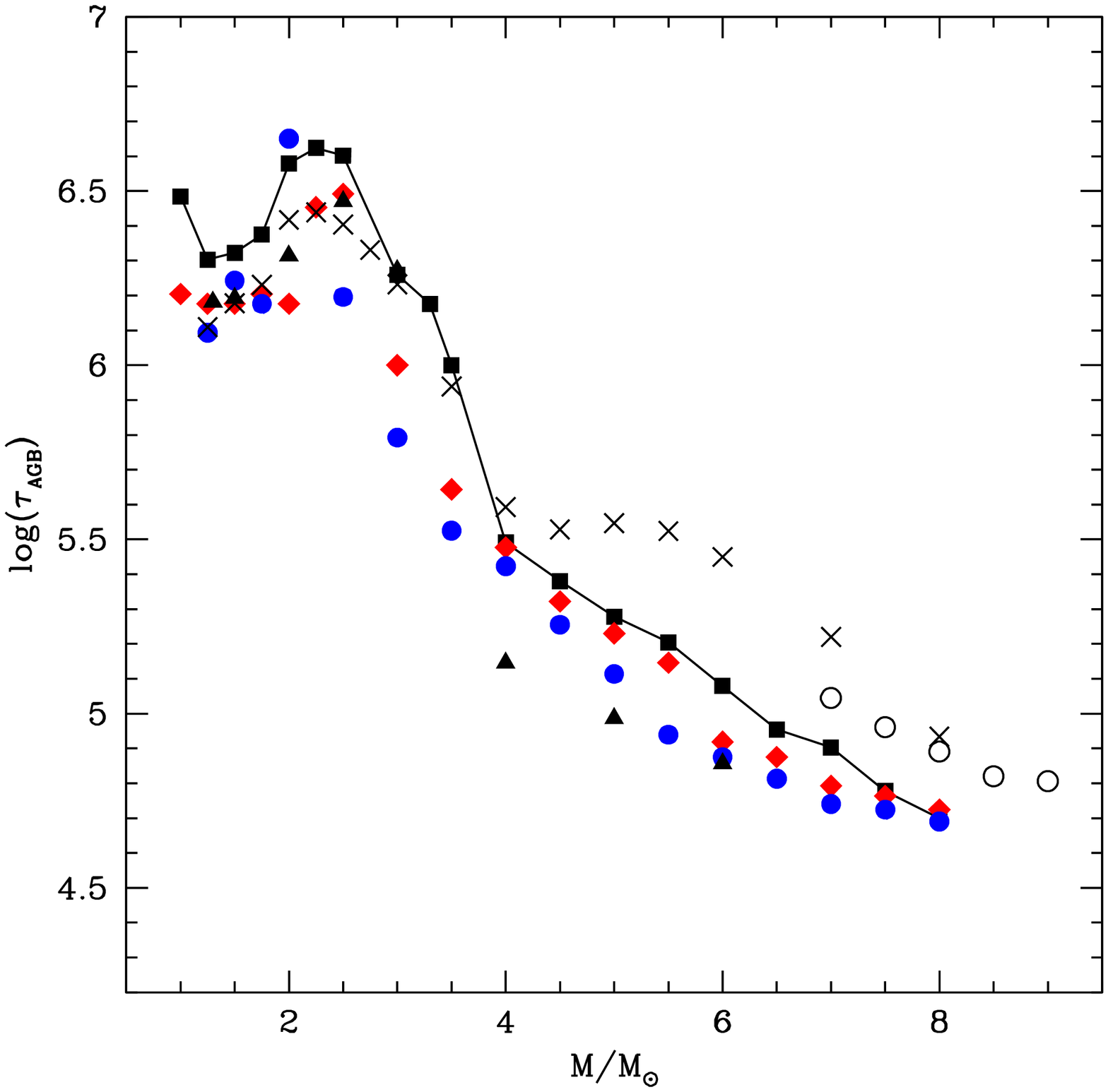}}
\end{minipage}
%\vskip-50pt
\caption{
Solar metallicity AGB model properties for various initial masses (full squares) are presented here.
The individual panels show the maximum luminosity reached (top, left), the highest temperature 
experienced at the base of the external mantle (top, right), the duration of the
TP-AGB phase (bottom, right) and the core masses at the beginning and at the end
of the AGB evolution (bottom, left).
The models at $Z=4\times 10^{-3}$
and $Z=8\times 10^{-3}$ metallicities are indicated, respectively, with blue full circles and
red diamonds. For comparison, we also show the results from \citet{cristallo09, cristallo15} 
(triangles), \citet{karakas16} (crosses) and \citet{doherty14} (open circles).
}
\label{fagb}
\end{figure*}

The evolution of stars of low- and intermediate mass through the AGB phase is mainly
driven by the mass of the degenerate core, which determines the brightness of the star,
the time required to lose the external mantle and the relative importance of the two
mechanisms potentially able to alter the surface chemical composition, namely HBB
and TDU. Exhaustive reviews, with detailed explanations of the most important properties
of stars evolving through the asymptotic giant branch and the uncertainties related to
their description, were published by \citet{herwig05} and \citet{karakas14b}.

A summary of the main physical properties of the models presented here is reported in
Table 1 and in Fig.~\ref{fagb}, showing the duration of the 
%% nella caption c'e' TP-AGB? anche l'ordine e' diverso
AGB phase, the maximum
luminosity experienced ($L_{max}$), the core mass at the beginning and at the end of 
the AGB phase and the largest temperature reached at the base of the convective envelope
($T_{bce}^{max}$). In the same figure we also show the results of lower metallicity models 
\citep{ventura13, ventura14a, ventura14b}, and solar metallicity models 
calculated by other research groups \citep{karakas16, doherty14, cristallo15}.

All the physical quantities show clear trend with the initial mass ($M_{init}$); an upturn 
in the core mass vs. $M_{init}$ relationship is found around $\sim 2~\Msun$, 
at the transition between lower mass stars, undergoing the helium flash, and more massive
objects, experiencing core helium burning  ignition in conditions of thermal stability. \\
Both $L_{max}$  vs $M_{init}$ and  $T_{bce}^{max}$ vs $M_{init}$ trend outlines an abrupt transition occurring for 
masses slightly above $3~\Msun$, consequently to the ignition of HBB. As thoroughly
documented in the literature \citep{vd05a}, the occurrence of HBB has a significant 
impact on the AGB evolution. Stars undergoing HBB evolve to brighter luminosities
\citep{blocker91} and experience a fast loss of their external mantle; 
on the chemical side, the surface
composition reflects the outcome of the nucleosynthesis experienced at the bottom of
the surface convective region.
Based on these reasons, in the following we discuss separately the main properties of 
the stars experiencing HBB and the objects of mass below $3~\Msun$.

%%%%%%%%%%%%%%%%%%%%%%%%%%%%%%%%%%%%%%%%%%%%%%%%%%%%%%%%%%%
\subsection{Massive AGB stars}
Stars with initial mass $3~\Msun < M_{init} \leq 8~\Msun$ 
experience HBB at the base of the convective envelope\footnote{This mass range depends on 
metallicity, i.e. lower-Z stars achieve HBB conditions more easily . The lower
mass limit to experience HBB decreases to $\sim 2.5\Msun$ for metallicities below 
$Z=4\times 10^{-3}$.}. 
Within this mass interval we separate $3~\Msun < M_{init} < 6.5~\Msun$ stars (which develop a carbon-oxygen core) 
and $6.5~\Msun < M_{init} < 8~\Msun$ objects, which (after the carbon ignition in a partially degenerate off-center zone) 
develop an oxygen-neon core \citep{garcia94, garcia97, siess06, siess07, siess09, siess10}.

\begin{figure}
%\resizebox{1.\hsize}{!}{\includegraphics{figlumhbb.ps}}
\resizebox{1.\hsize}{!}{\includegraphics{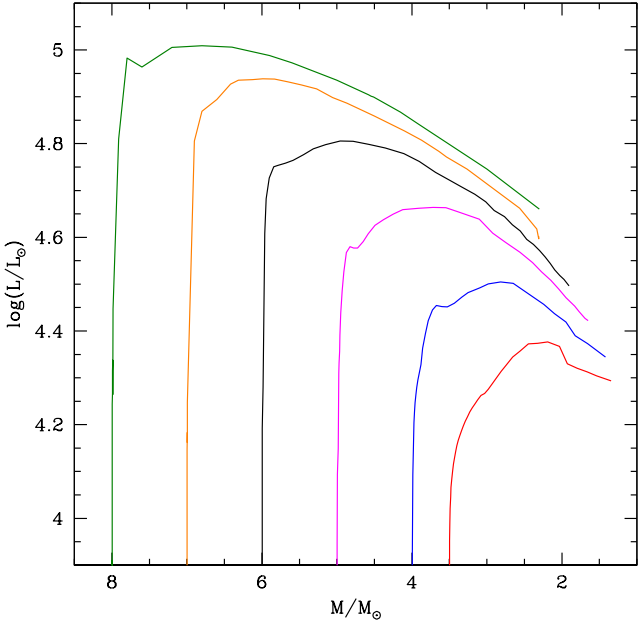}}
%\vskip-50pt
\caption{
The AGB evolution of the maximum surface luminosity reached by  stars of
different mass, experiencing HBB.
On the abscissa we report the mass of the star (decreasing during the 
evolution). The various tracks correspond to model of initial mass $3.5~\Msun$ (red),
$4~\Msun$ (blue), $5~\Msun$ (magenta), $6~\Msun$ (black), $7~\Msun$ (orange), 
$8~\Msun$ (green).
}
\label{flum}
\end{figure}

%%%% refereee quindi non ho corretto per ora
 On general grounds, the maximum luminosity reached by stars undergoing HBB evolves to brighter and brighter luminosities during the initial AGB phases, as a consequence
of the increase in the core mass; in more advanced
phases the overall luminosity declines, owing to the gradual loss of the external mantle,
which provokes a general cooling of the whole external zones, that reduces the
efficiency of the CNO activity. This behaviour can be seen in Fig.~\ref{flum}, showing the
AGB evolution of the surface luminosity of models of different initial mass; we used
the (current) mass of the star as abscissa, to allow the simultaneous plot of all the models.
As clear from Fig.~\ref{flum} (see also top, left panel of Fig.~\ref{fagb}) the
highest luminosity experienced is extremely sensitive to $M_{init}$, ranging
from $\sim 25000~L_{\odot}$ for the $M_{init}=3.5~\Msun$ model, to $\sim 10^5~L_{\odot}$ 
for $M_{init}=8~\Msun$.\\
The luminosity dependency on initial mass is determined by the larger core masses of larger initial mass models, 
as shown in the left bottom panel of Fig.~\ref{fagb}. Core masses range from $\sim 0.7~\Msun$
($M_{init} = 3.5~\Msun$) to $\sim 1.25~\Msun$ ($M_{init} = 8~\Msun$). 
Higher initial mass models experience a faster loss of the external envelop and thus a shorter AGB phase, 
because larger luminosities imply larger mass loss rates.
While the AGB phase of a $3.5~\Msun$ star lasts $\sim 10^6$ yr, 
in the case of the $8~\Msun$ star it is limited to $\sim 5\times 10^4$ yr (see right, 
bottom panel of Fig.~\ref{fagb}).\footnote{A word of caution is needed here: the short duration of 
the AGB phase of massive AGBs, particularly of the stars whose initial mass is close to 
the threshold limit to undergo core collapse, is partly due to the steep dependence on luminosity
of the mass loss rate used here \citep{blocker95} ; the interested reader
can find in \citet{doherty14} an exhaustive discussion on this subject.} 

The core mass also affects the temperature at the base of the convective envelope,
which, as shown in the right,top panel of Fig.~\ref{fagb}, increases linearly with
mass, ranging from $\sim 60$ MK ($M_{init} = 3.5~\Msun$) to $\sim 110$ MK 
($M_{init} = 8~\Msun$). Models of higher mass are therefore expected to experience
a stronger HBB, with a more advanced nucleosynthesis at the base of the convective
envelope.

Fig.~\ref{fagb} allows to appreciate the effects of metallicity: lower metallicity models
reach higher temperatures at the base of the envelope, thus they experience stronger HBB
conditions, and their external regions are exposed to a more advanced nucleosynthesis.

The surface chemical composition of massive AGB stars is mainly determined by HBB, with a 
modest contribution from TDU. The effects of the latter mechanism are more evident
towards the latest evolutionary phases, when HBB is turned off by the gradual consumption
of the external envelope. In stars of mass around $\sim 3.5 \Msun$, with an initial mass
just above the threshold necessary to activate HBB, the evolution of the surface
chemistry is given by the balance of the two mechanisms.

\begin{figure*}
\begin{minipage}{0.33\textwidth}
\resizebox{1.\hsize}{!}{\includegraphics{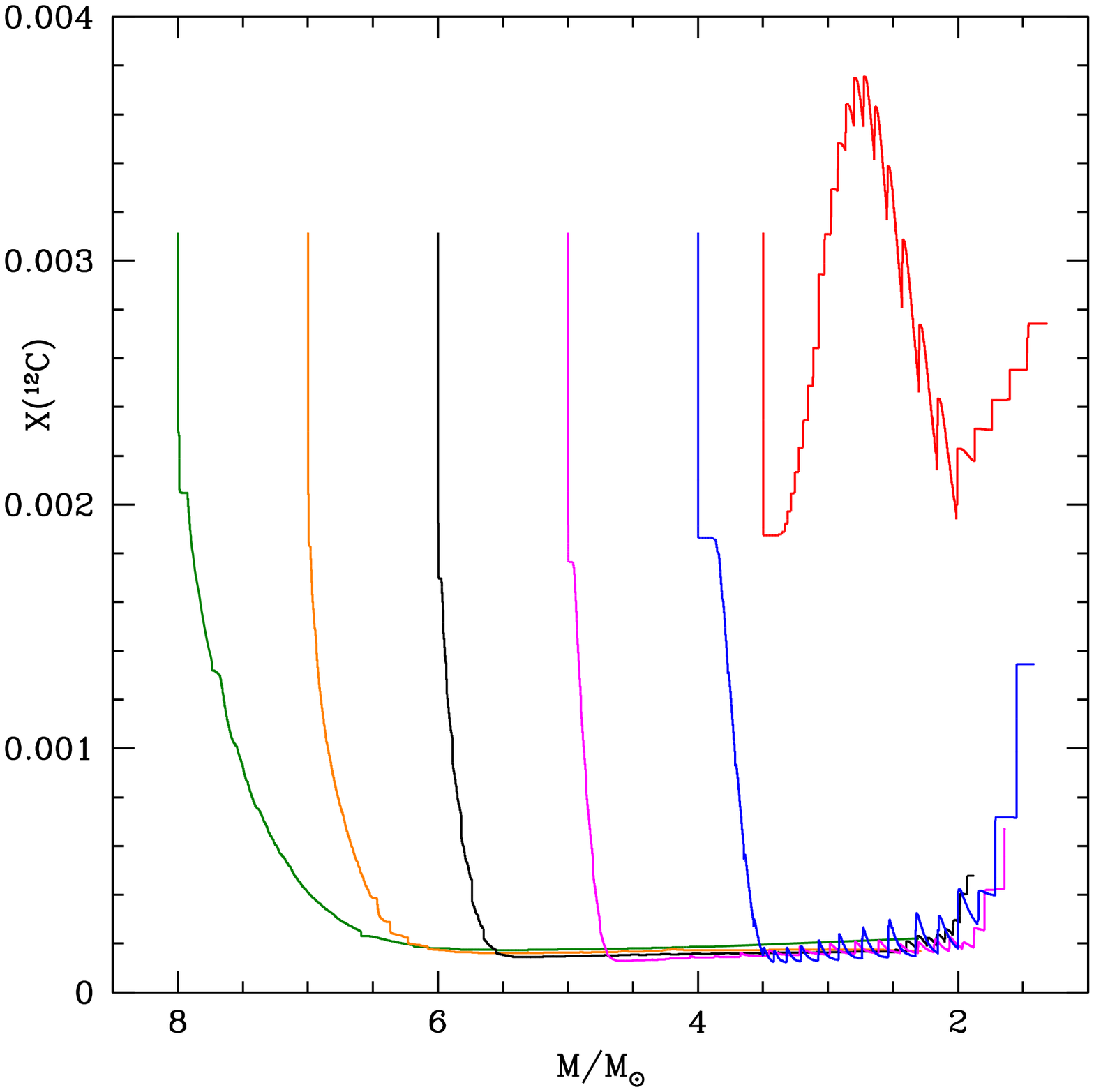}}
%\resizebox{1.\hsize}{!}{\includegraphics{figc12hbb.ps}}
\end{minipage}
\begin{minipage}{0.33\textwidth}
%\resizebox{1.\hsize}{!}{\includegraphics{fign14hbb.ps}}
\resizebox{1.\hsize}{!}{\includegraphics{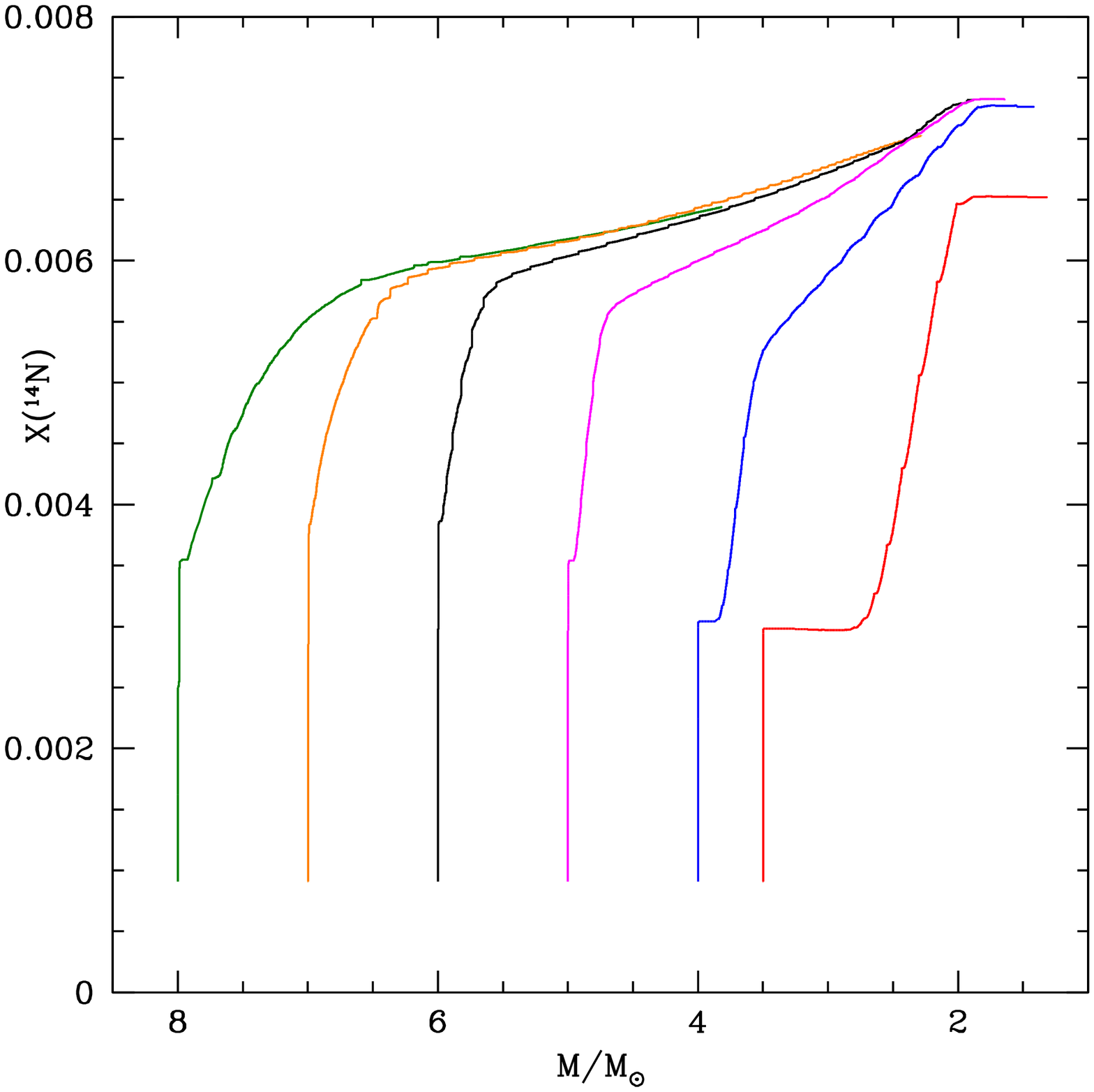}}
\end{minipage}
\begin{minipage}{0.33\textwidth}
%\resizebox{1.\hsize}{!}{\includegraphics{figo16hbb.ps}}
\resizebox{1.\hsize}{!}{\includegraphics{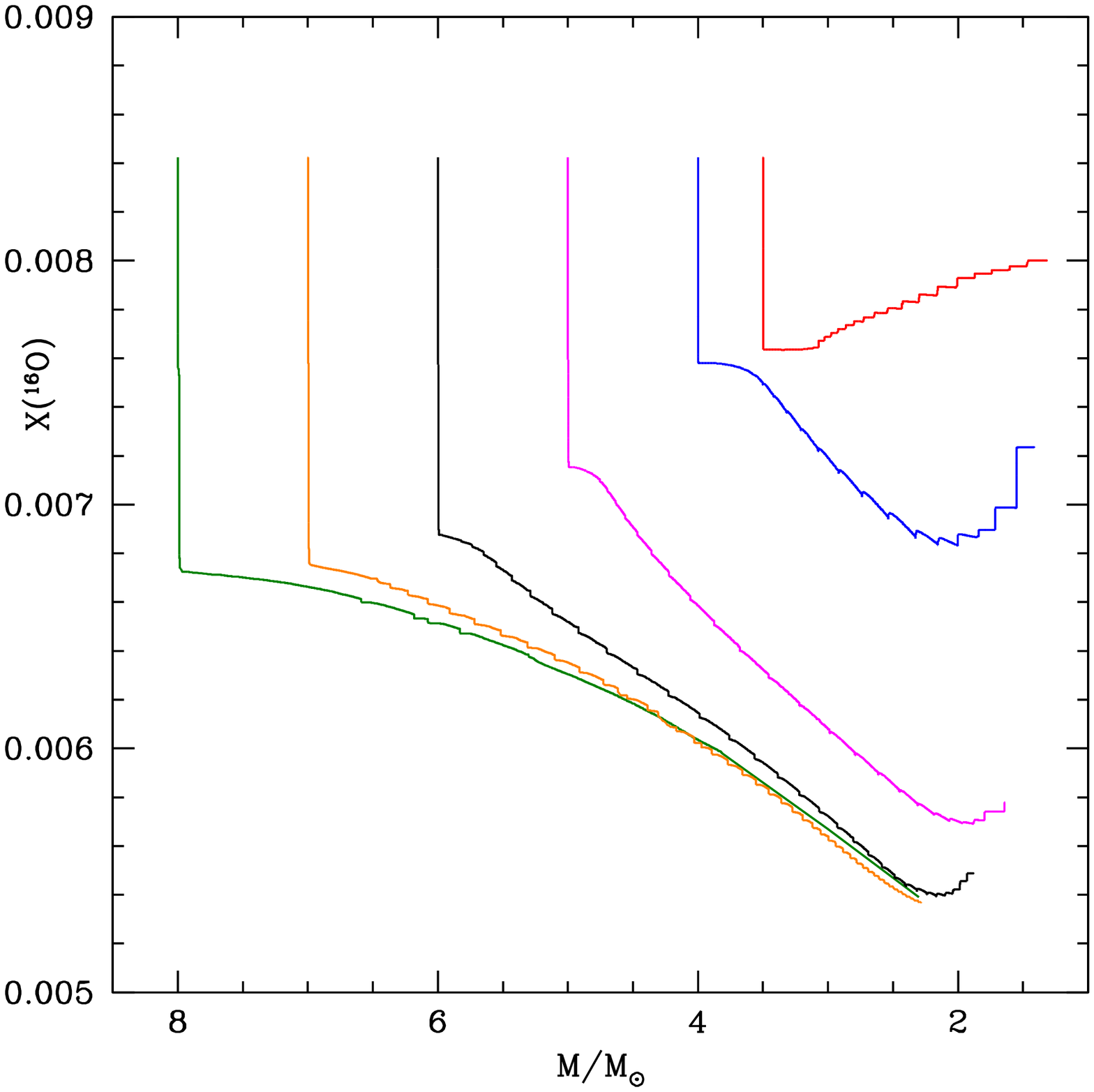}}
\end{minipage}
%\vskip-30pt
\caption{The variation of the surface abundance of $^{12}C$ (left panel), 
$^{14}N$ (middle), $^{16}O$ (right) during the same AGB models shown in 
Fig.~\ref{flum}.
}
\label{fcno}
\end{figure*}

\subsubsection{CNO cycling} 
Fig.~\ref{fcno} shows the variation with time of the CNO elements surface mass fraction
in stars experiencing HBB. The surface carbon diminishes by $\sim 30\%$ during the
first dredge-up episode and is further destroyed during the AGB phase,  since the early
TPs. Independently of the initial mass, an equilibrium is reached, where
the surface carbon is $\sim 50$ smaller than the initial value and the $^{12}C/^{13}C$
ratio is $\sim 4$; as clearly shown in the figure, most of the mass ejected by these stars 
has this chemical composition. 
In the final AGB phases, when HBB is no longer active, some carbon is transported to the surface by TDU; this is
particularly evident in the tracks corresponding to $4 \Msun$ and $4.5 \Msun$ models.
The $3.5\Msun$ star follows a different behaviour, with the AGB evolution 
divided into three phases: a) the initial phase, when the surface carbon increases owing to 
the effects of TDU; b) an intermediate phase, when HBB destroys the carbon previously accumulated; 
c) the final TPs, when HBB is turned off and the surface carbon increases again.

\begin{table*}
\setlength{\tabcolsep}{3pt}
\caption{Chemical yields of solar metallicity models}                                       
\begin{tabular}{c c c c c c c c c c c c c c c}        
\hline\hline                        
M & He & $^{12}C$ & $^{13}C$ & $^{14}N$ & $^{15}N$ & $^{16}O$ & $^{17}O$ & 
$^{18}O$ & $^{22}Ne$ & $^{23}Na$ & $^{24}Mg$ & $^{25}Mg$ & $^{26}Mg$ & $^{27}Al$  \\
\hline       
1.25 & 1.55(-2) &  1.15(-4)  & 3.91(-5) & 4.37(-4) & -1.91(-8) &  3.70(-5) & 2.29(-6) & -1.15(-6) &  2.85(-5) & 3.59(-7) &   0 & 0 & 0 & 0 \\
1.5  & 2.06(-2) &  1.77(-3)  & 4.81(-5) & 8.22(-4) & -3.25(-8) &  1.47(-4) & 2.59(-6) & -2.82(-6) &  1.50(-4) & 2.21(-6) & -1.46(-7) & 2.08(-6) & 1.90(-6) & 4.86(-8) \\
1.75 & 2.75(-2) &  3.26(-3)  & 5.93(-5) & 1.28(-3) & -4.61(-8) &  2.55(-4) & 1.24(-5) & -4.69(-6) &  2.76(-4) & 4.07(-6) & -3.66(-7) & 4.66(-6) & 3.98(-6) & 1.44(-7) \\
2.0  & 2.00(-2) &  4.13(-3)  & 7.30(-5) & 2.02(-3) & -5.56(-8) &  3.57(-4) & 2.31(-5  & -6.53(-7) &  3.53(-4) & 4.94(-6) & -4.89(-7) & 6.15(-6) & 5.18(-6) & 2.23(-7) \\
2.25 & 4.41(-2) &  8.23(-3)  & 8.65(-5) & 2.64(-3) & -6.31(-8) & -9.87(-5) & 3.50(-5) & -8.19(-6) &  7.07(-4) & 1.08(-5) & -1.54(-6) & 1.88(-5) & 1.37(-5) & 1.16(-6) \\
2.5  & 6.56(-2) &  1.09(-2)  & 8.91(-5) & 2.49(-3) & -6.67(-8) & -1.26(-4) & 2.66(-5) & -8.36(-6) &  9.11(-4) & 1.45(-5) & -1.68(-6) & 2.18(-5) & 1.63(-5) & 1.21(-6) \\
3.0  & 7.87(-2) &  1.38(-2)  & 1.27(-4) & 4.54(-3) & -8.56(-8) & -3.94(-4) & 2.92(-5) & -1.25(-5) &  1.28(-3) & 2.18(-5) & -7.27(-6) & 7.71(-5) & 4.42(-5) & 1.09(-5) \\
3.5  & 8.09(-2) & -1.16(-3)  & 1.15(-3) & 1.11(-2) & -2.14(-7) & -1.55(-3) & 2.40(-5) & -3.77(-5) &  4.00(-4) & 9.00(-5) & -1.41(-5) & 6.29(-5) & 3.88(-5) & 3.79(-5) \\
4.0  & 9.04(-2) & -7.80(-3)  & 1.25(-4) & 1.67(-2) & -2.88(-7) & -3.93(-3) & 3.88(-5) & -5.32(-5) & -2.73(-4) & 4.48(-4) & -1.91(-5) & 4.36(-5) & 1.96(-5) & 3.83(-5) \\
4.5  & 1.72(-1) & -1.01(-2)  & 7.60(-5) & 1.99(-2) & -3.36(-7) & -6.82(-3) & 4.91(-5) & -6.25(-5) & -3.70(-4) & 4.74(-4) & -5.39(-5) & 4.20(-5) & 1.58(-5) & 2.78(-5) \\
5.0  & 2.59(-1) & -1.18(-2)  & 6.82(-5) & 2.30(-2) & -4.01(-7) & -9.15(-3) & 6.07(-5) & -7.13(-5) & -4.51(-4) & 5.13(-4) & -1.14(-4) & 8.48(-5) & 2.91(-5) & 1.27(-5) \\
5.5  & 3.48(-1) & -1.31(-2)  & 8.75(-5) & 2.61(-2) & -4.40(-7) & -1.10(-2) & 7.25(-5) & -7.98(-5) & -5.03(-4) & 5.49(-4) & -2.00(-4) & 1.63(-4) & 3.97(-5) & 1.49(-5) \\
6.0  & 4.33(-1) & -1.42(-2)  & 8.71(-5) & 2.83(-2) & -4.81(-7) & -1.22(-2) & 8.92(-5) & -8.80(-5) & -5.50(-4) & 5.80(-4) & -3.05(-4) & 2.72(-4) & 4.95(-5) & 1.88(-5) \\
6.5  & 5.17(-1) & -1.60(-2)  & 8.65(-5) & 3.07(-2) & -5.62(-7) & -1.35(-2) & 1.09(-4) & -9.62(-5) & -6.11(-4) & 6.11(-4) & -4.66(-4) & 3.96(-4) & 5.62(-5) & 1.07(-5) \\
7.0  & 5.67(-1) & -1.71(-2)  & 1.24(-4) & 3.27(-2) & -6.54(-7) & -1.45(-2) & 1.33(-4) & -1.04(-4) & -6.48(-4) & 6.24(-4) & -7.76(-4) & 7.11(-4) & 3.99(-5) & 4.98(-5) \\
7.5  & 6.02(-1) & -1.79(-2)  & 2.13(-4) & 3.41(-2) & -8.02(-7) & -1.49(-2) & 1.82(-4) & -1.11(-4) & -6.63(-4) & 6.42(-4) & -9.12(-4) & 9.22(-4) & 4.03(-5) & 5.12(-5) \\
8.0  & 6.34(-1) & -1.89(-2)  & 2.66(-4) & 3.52(-2) & -9.17(-7) & -1.52(-2) & 2.06(-4) & -1.17(-4) & -6.88(-4) & 6.60(-4) & -1.15(-3) & 1.13(-3) & 4.10(-5) & 5.59(-5) \\
\hline     
\end{tabular}
\end{table*}

The destruction of the surface carbon is related 
to the relatively low temperatures required to activate carbon burning at the base of the 
envelope of AGB stars, namely $T_{bce} \sim40$ MK; as shown in 
Fig.~\ref{fagb}, these $T_{bce}$'s are reached by all models experiencing HBB during the 
initial AGB phase. The only exception is the $3.5\Msun$ model, where the temperature 
necessary to start proton capture nucleosynthesis by $^{12}C$ nuclei is reached in more 
advanced AGB phases, after some TDU episodes occurred (see left panel of Fig.~\ref{fcno}).

The activation of the CNO nucleosynthesis leads to the synthesis of nitrogen, which is
increased (see middle panel of Fig.~\ref{fcno}) almost by an order of magnitude at the
surface of the stars. It is worth noticing that most of this nitrogen has a secondary origin in the present
models, as N is essentially produced by the carbon originally present in the star. %%???

The evolution of surface oxygen abundance is more complicated, as the activation of the whole CNO
cycle (with the oxygen destruction) requires temperatures significantly higher than
those necessary for the carbon burning ignition, namely $\sim 80$ MK. This makes
oxygen depletion extremely sensitive to  mass and
chemical composition, as these are the two most relevant quantities
in the determination of the temperature at which HBB occurs. \citet{ventura13} showed
that massive AGBs at $Z=3\times 10^{-4}$ metallicity produce ejecta with
an oxygen content a factor 10 smaller compared to the gas from which the stars formed.
On the contrary, higher metallicity AGBs ($Z=8\times 10^{-3}$) were shown to undergo a less
advanced nucleosynthesis and to eject gas with an oxygen content on average
$\sim 0.2$ smaller than the initial value. \\
As discussed earlier in this section (see also top right panel of Fig.~\ref{fagb}), 
solar metallicity models have a less efficient HBB compared 
to lower metallicity models. Therefore, the surface oxygen 
survives more easily in the solar metallicity models. 
As shown in Fig.~\ref{fcno}, the lowest oxygen abundances ($\sim 30-40 \%$ below the initial values), are 
present in the most massive models evolution, in the final AGB phases. 
%%% ricontrollare alle volte 0.2 alle volte 30% scrivere o frazione o percentuale senno' si segue male
%% se si puo'
For $M_{init} \leq 4\Msun$ the surface oxygen decreases during the second dredge-up event and is produced during the 
following AGB phase, owing to the effects of TDU.\\
Considering oxygen isotopes, the HBB nucleosynthesis is accompanied by 
a considerable destruction of the surface $^{18}O$, which is rapidly consumed starting with 
the early TPs, when it reaches an equilibrium abundance of $^{18}O/^{16}O \sim 10^{-6}$. 
%%% mi sono persa rispetto a quello che scrivi sopra sulla temperatura per distruzione ossigeno....
The destruction of $^{18}O$ occurs at the same temperatures required for carbon burning ignition. 
On the contrary, $^{17}O$ is produced as soon as
$^{16}O$ burning begins, the overall production factor ranging from 5 to 10, depending on
the initial mass of the star. The variation of the $^{18}O/^{17}O$ ratio of the models discussed here is shown in the right panel of Fig.11.

%%%%%%%%%%%%%%%%%%%%%%%%%%%%%%%%%%%%%%%%%%%%%%%%%%%%%%%%%%%%%
\subsubsection{Sodium production}
The Ne-Na nucleosynthesis is activated at the same temperatures at which oxygen burning occurs. 
The evolution of surface sodium abundance during the AGB phase is complicated
\citep{mowlavi99} and depends on the balance between the production channel (i.e. 
the proton capture process by $^{22}Ne$ nuclei) and the destruction reactions ($^{23}Na(p,\gamma)^{24}Mg$ and $^{23}Na(p,\alpha)^{20}Ne$ 
reactions, with the latter providing the dominant contribution). 
The production mechanisms prevail at temperatures lower than 90MK, whereas the destruction reactions, 
whose cross sections have a steeper dependance on temperature, become dominant for T$>$90 MK. 
At the beginning of the AGB phase sodium is thus produced via $^{22}Ne$ burning, whereas it is destroyed in more 
advanced phases, when the destruction processes predominate \citep{ventura06, ventura08}. 

\begin{figure}
%\resizebox{1.\hsize}{!}{\includegraphics{figsodiohbb.ps}}
\resizebox{1.\hsize}{!}{\includegraphics{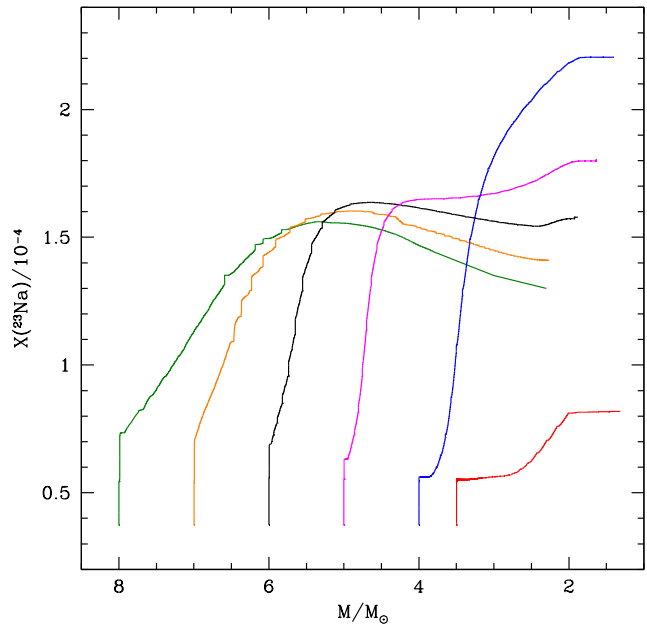}}
%\vskip-50pt
\caption{The variation of the surface sodium mass fraction (in $10^{-4}$ units) of 
AGB models experiencing HBB. The colour coding is the same as in Fig.~\ref{flum}.
}
\label{fsodio}
\end{figure}

The variation of the sodium surface abundance during the AGB phase for the solar metallicity models is shown in 
Fig.~\ref{fsodio}. The dependency on initial mass offers an interesting example of how the
temperature at the base of the convective zone is crucial to determine the
nucleosynthesis in these stars. The $T_{bce}$  dependency on $M_{init}$ shown in the top
right panel of Fig.~\ref{fagb}, explains the results of Fig.~\ref{fsodio}.  
In stars with $M_{init} \geq 6~\Msun$, sodium is produced in the initial AGB phases and partly destroyed later on, when 
$T_{bce}$ exceeds 90 MK and HBB reaches the strongest efficiency. In stars with
$4~\Msun \leq M_{init} < 6~\Msun$ sodium is produced during the whole AGB
phase, with no destruction, because the temperature at the base of the external mantle
is lower than 90 MK (see top, right panel of Fig. 1 and Table 1). For stars with initial mass 
just above the threshold to activate HBB (here represented by the $3.5~\Msun$ star), only a small production of sodium occurs, 
because the temperature is not high enough to allow an efficient $^{22}Ne$ burning.

In summary, unlike the stars of lower metallicity \citep{ventura11}, here the 
destruction processes never really predominate, because of the lower temperatures
reached at the bottom of the convective envelope. This results into a significant 
increase in the surface sodium, with final abundances 4-5 times larger than the
initial values. The highest sodium production is reached in the $4\Msun$ model,
because the destruction reactions are never activated during the entire AGB life.

%%%%%%%%%%%%%%%%%%%%%%%%%%%%%%%%%%%%%%%%%%%%% 
\subsubsection{Mg-Al nucleosynthesis}
The magnesium-aluminum nucleosynthesis is activated at HBB temperatures close to 100MK: 
the proton capture by $^{24}Mg$ nuclei starts a series of reaction, whose outcome is the increase in the
surface content of the two heavier isotopes of magnesium and the aluminum synthesis \citep{arnould99, siess08}. 
\citet{ventura13} describes the extreme sensitivity of the Mg-Al nucleosynthesis efficiency to metallicity. 
As consequence of the different HBB strength at different chemical composition,
in low metallicity stars a significant production of aluminum occurs, whereas in objects with higher 
metallicities magnesium burning is less efficient, with a more limited aluminum synthesis. 

In the present solar metallicity models the activation 
the Mg-Al nucleosynthesis is limited to stars with $M_{init} \geq 5\Msun$.
The largest $^{24}Mg$ depletion ($\delta \log(^{24}Mg) \sim -0.15$ dex) is found in the
largest initial mass models; in all cases no significant aluminum synthesis occurs.

\begin{figure*}
\begin{minipage}{0.49\textwidth}
%\resizebox{1.\hsize}{!}{\includegraphics{figlitiohbb.ps}}
\resizebox{1.\hsize}{!}{\includegraphics{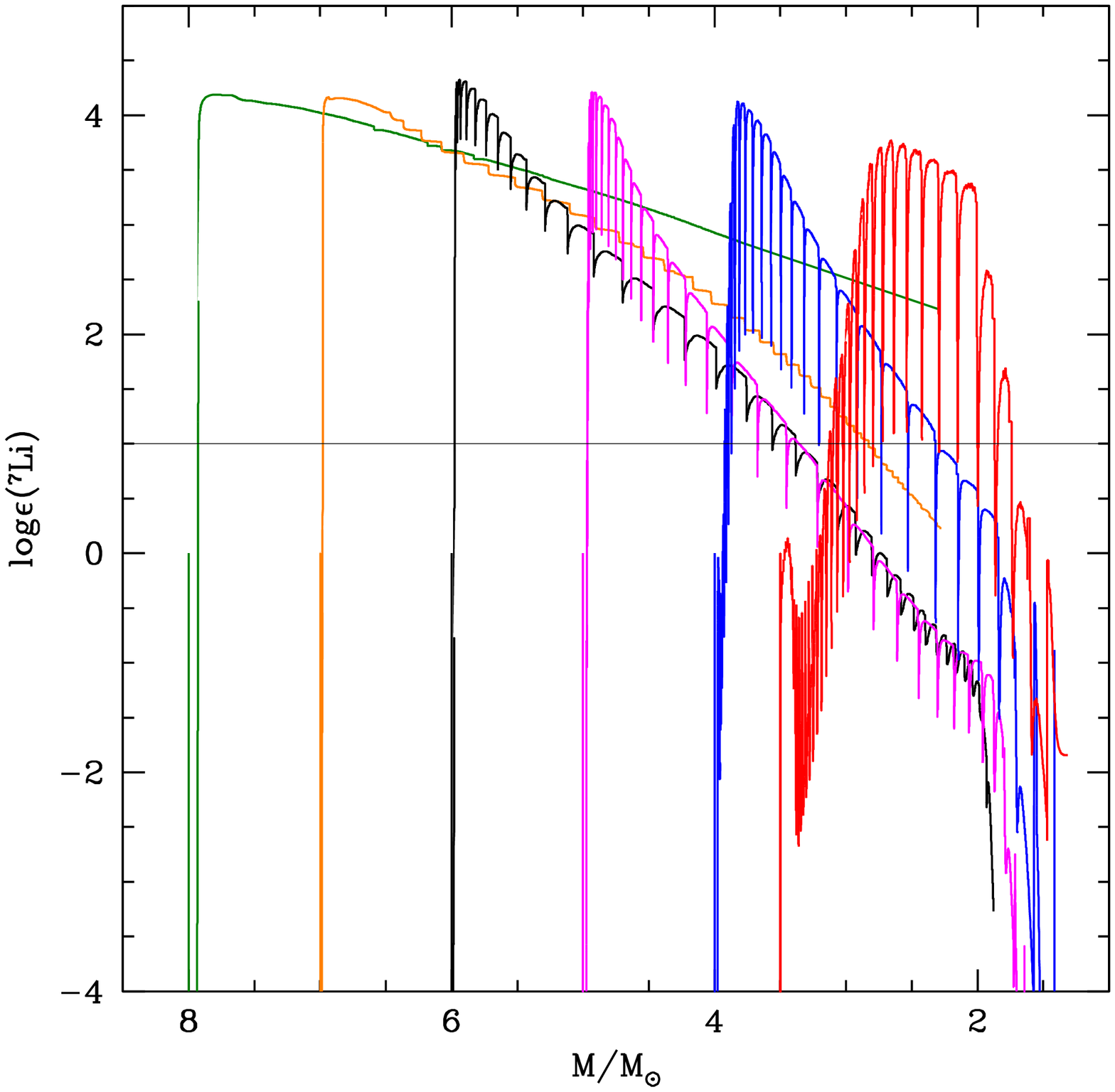}}
\end{minipage}
\begin{minipage}{0.49\textwidth}
%\resizebox{1.\hsize}{!}{\includegraphics{figlitiotime.ps}}
\resizebox{1.\hsize}{!}{\includegraphics{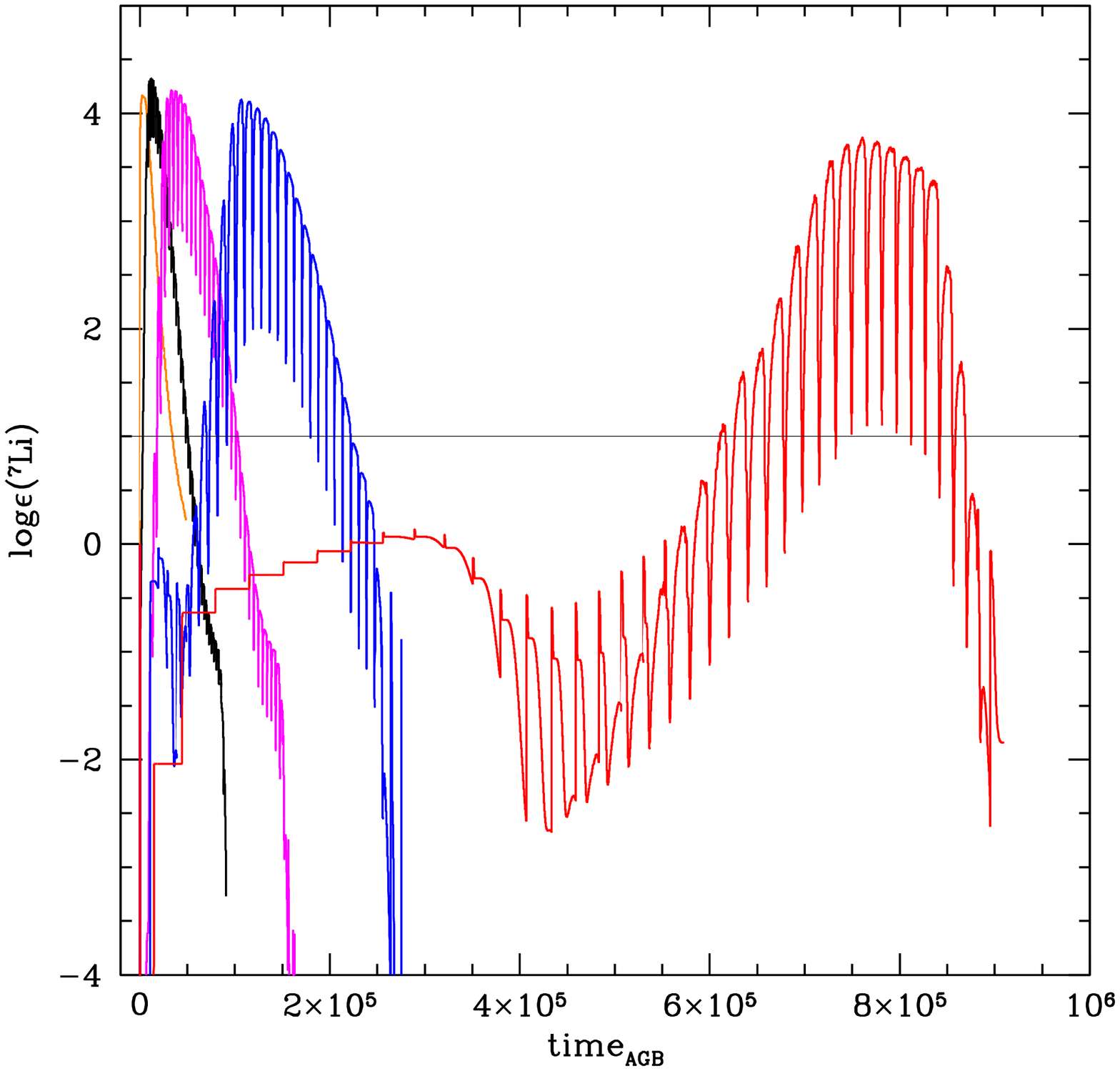}}
\end{minipage}
%\vskip-50pt
\caption{Surface lithium abundance evolution for the same models shown in
Fig.~\ref{flum}; the same colour coding is adopted. The quantity on the ordinate
is $\log(\epsilon (^7Li) = \log(^7Li/H)+12$. In the left panel we show the surface
lithium as a function of the initial mass, whereas on the right we use the AGB time
as abscissa. The horizontal line at $\log(\epsilon (^7Li) = 1$ indicates the limit
above which the stars are considered lithium-rich.
}
\label{flitio}
\end{figure*}

%%%%%%%%%%%%%%%%%%%%%%%%%%%%%%%%%%%%%%%%%%%%%%%%%%%%%
\subsubsection{Lithium}
\label{litio}

Lithium is synthetised during the AGB phase via the Cameron-Fowler
mechanism, which is started by the activation of $\alpha$ capture reactions by $^3He$ nuclei
at the base of the surface convective zone \citep{cameron71}. 
\citet{sackmann92} showed that the use of a self-consistent coupling between nuclear burning at the base of 
the envelope and mixing of chemicals in the same region, leads to production of great
quantities of lithium in the surface layers of AGB stars, provided that a minimum 
temperature of $\sim 30$ MK is reached at the base of the external mantle. As shown
in Fig.~\ref{fagb} and reported in Table 1, 
this property is shared by all the models presented here, with initial mass $M_{init} \geq 3.5~\Msun$, do reach the required
temperature.

Fig.~\ref{flitio} shows the variation of the surface lithium in our simulated  stars during the
AGB evolution: the results are shown as a function of the current mass of the stars and
of the time counted from the beginning of the AGB phase.

Lithium is produced since the early TP-AGB phases, as soon as HBB is activated. The
only exception to this  behaviour is the $3.5~\Msun$ model, in which lithium
production occurs in more advanced AGB phases, after the star has experienced a C-star 
phase. In agreement with the general understanding of the lithium production in these
objects, the surface lithium reaches a maximum abundance, after which it decreases 
below any detectability threshold. This apparently anomalous
behaviour (the temperature at the base of the envelope keeps increasing until after
the surface lithium is consumed, which would further favour the rate at which the
Cameron-Fowler mechanism works) is due to the exhaustion of the surface $^3He$ which is at the base of the nuclear chain leading to lithium production.

AGB stars of solar chemical composition are expected to have a longer lithium-rich phase compared to
metal poor AGBs because the smaller temperatures at the base of the envelope (see top, right panel of Fig.~\ref{fagb})
delay the surface $^3He$ exhaustion. 

As shown in the right panel of Fig.~\ref{flitio}, the lithium-rich phase lasts for
about half of the AGB evolution of these stars. The gas yields are therefore expected to show some lithium enrichment. 

Stars with initial mass higher than $\sim 7~\Msun$ are expected to be lithium-rich
for the whole AGB phase because their large mass loss rates make the time scale for
envelop consumption comparable to the $^3He$ destruction time scale.
This result must be taken with some caution though, as it is
strongly sensitive to the mass loss mechanism description.

\begin{figure*}
\begin{minipage}{0.48\textwidth}
%\resizebox{1.\hsize}{!}{\includegraphics{figlumcstar.ps}}
\resizebox{1.\hsize}{!}{\includegraphics{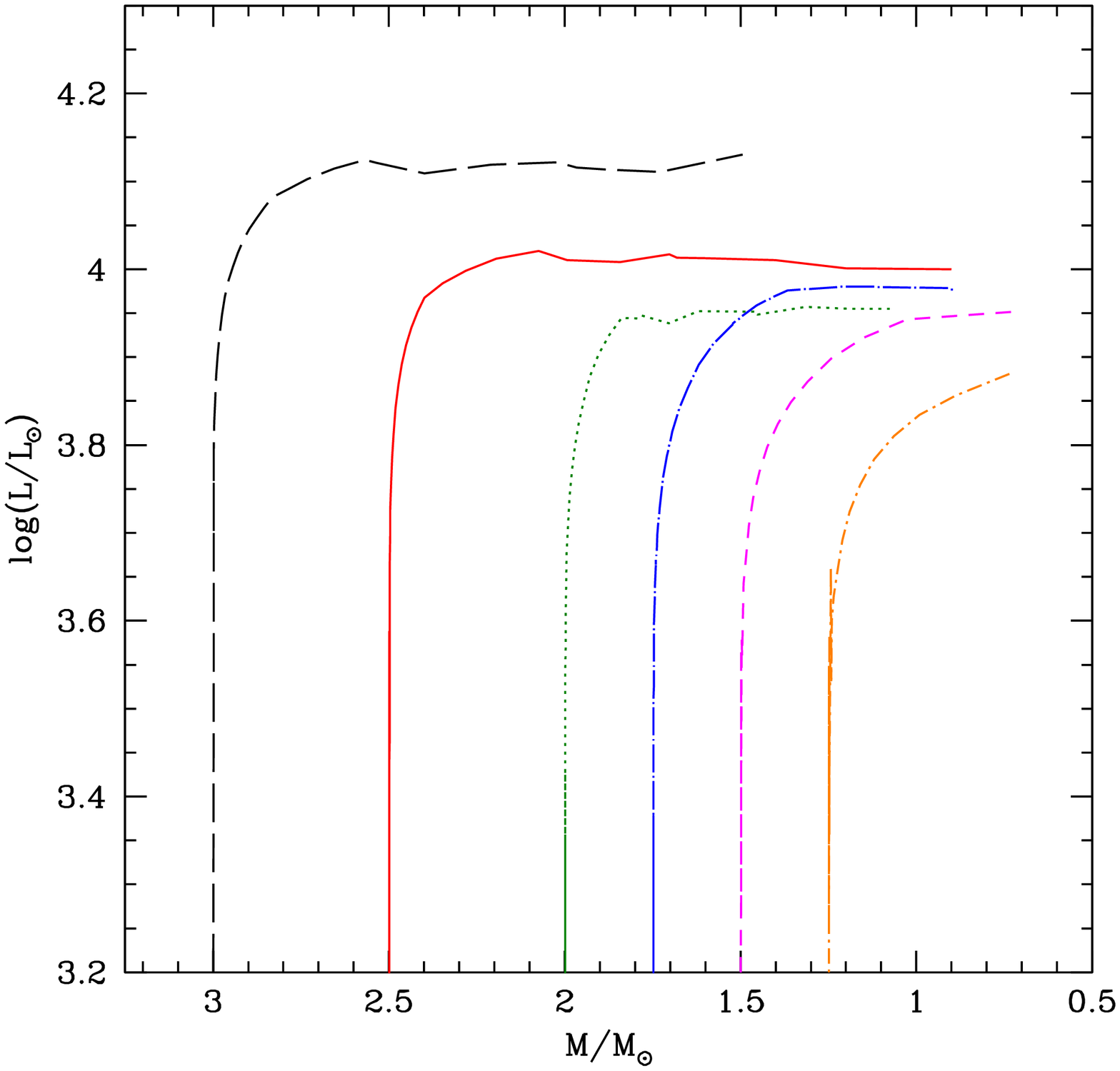}}
\end{minipage}
\begin{minipage}{0.48\textwidth}
%\resizebox{1.\hsize}{!}{\includegraphics{figtefcstar.ps}}
\resizebox{1.\hsize}{!}{\includegraphics{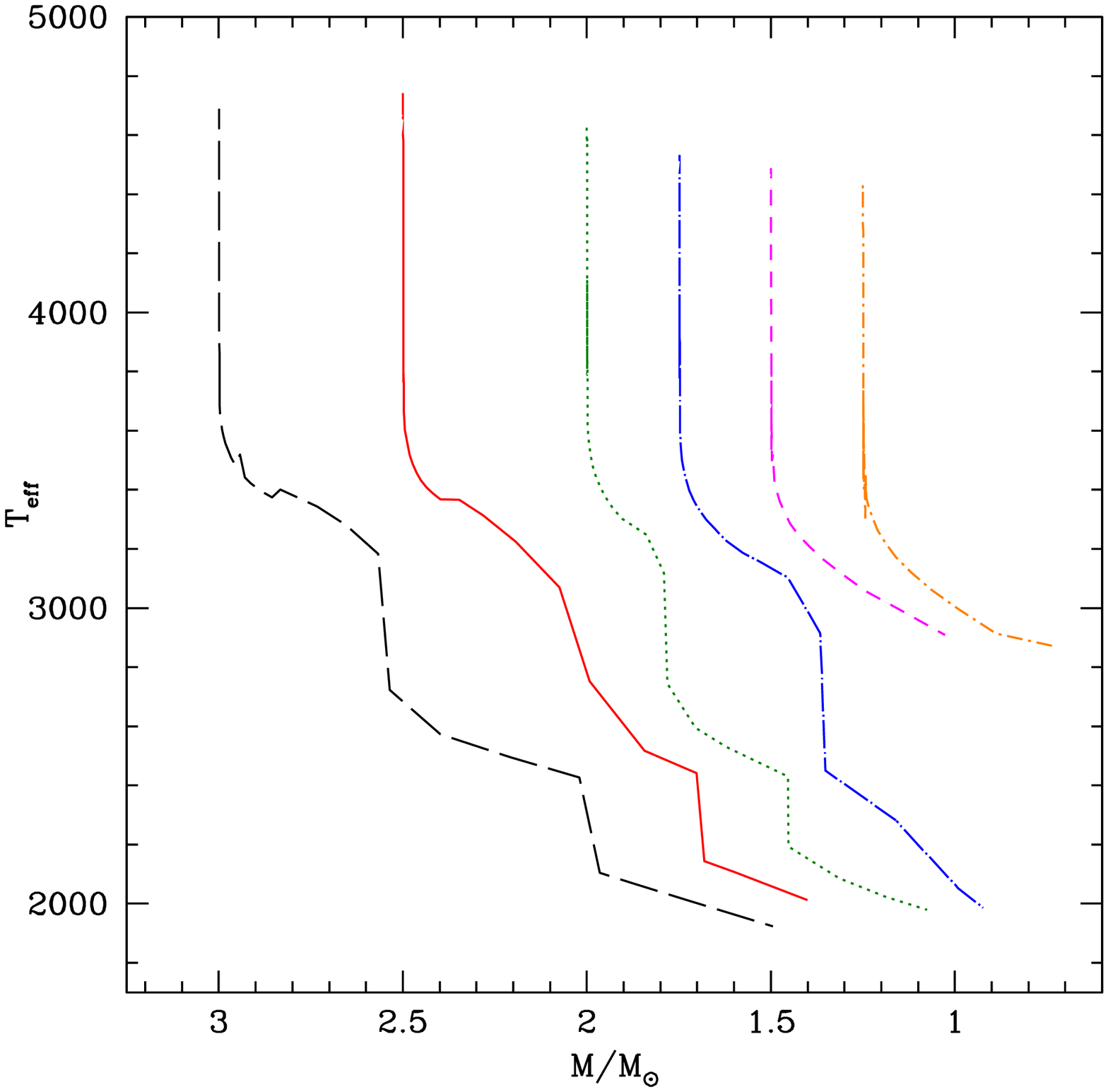}}
\end{minipage}
%\vskip-80pt
\begin{minipage}{0.48\textwidth}
%\resizebox{1.\hsize}{!}{\includegraphics{figmlosscstar.ps}}
\resizebox{1.\hsize}{!}{\includegraphics{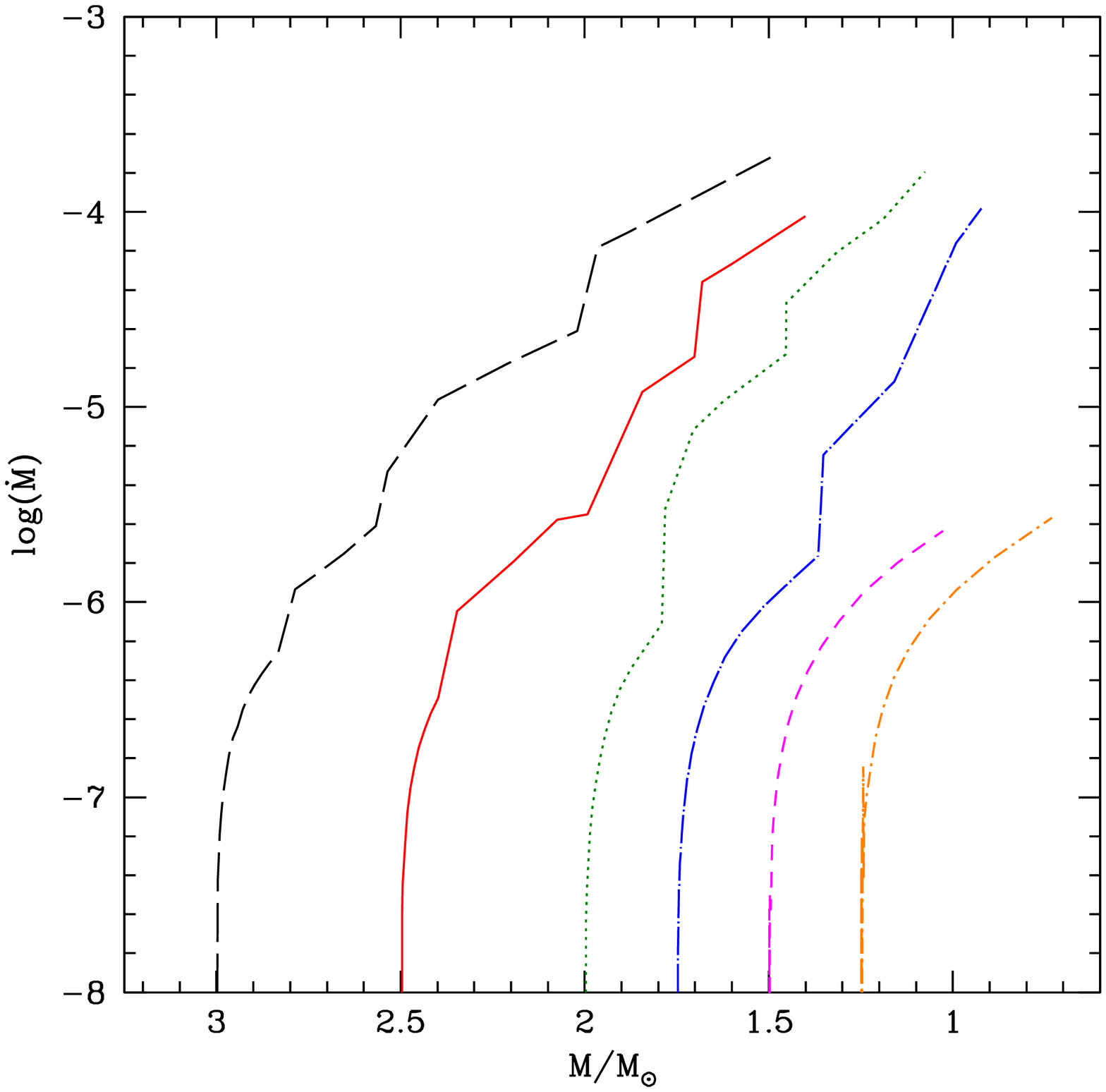}}
\end{minipage}
\begin{minipage}{0.48\textwidth}
%\resizebox{1.\hsize}{!}{\includegraphics{figcocstar.ps}}
\resizebox{1.\hsize}{!}{\includegraphics{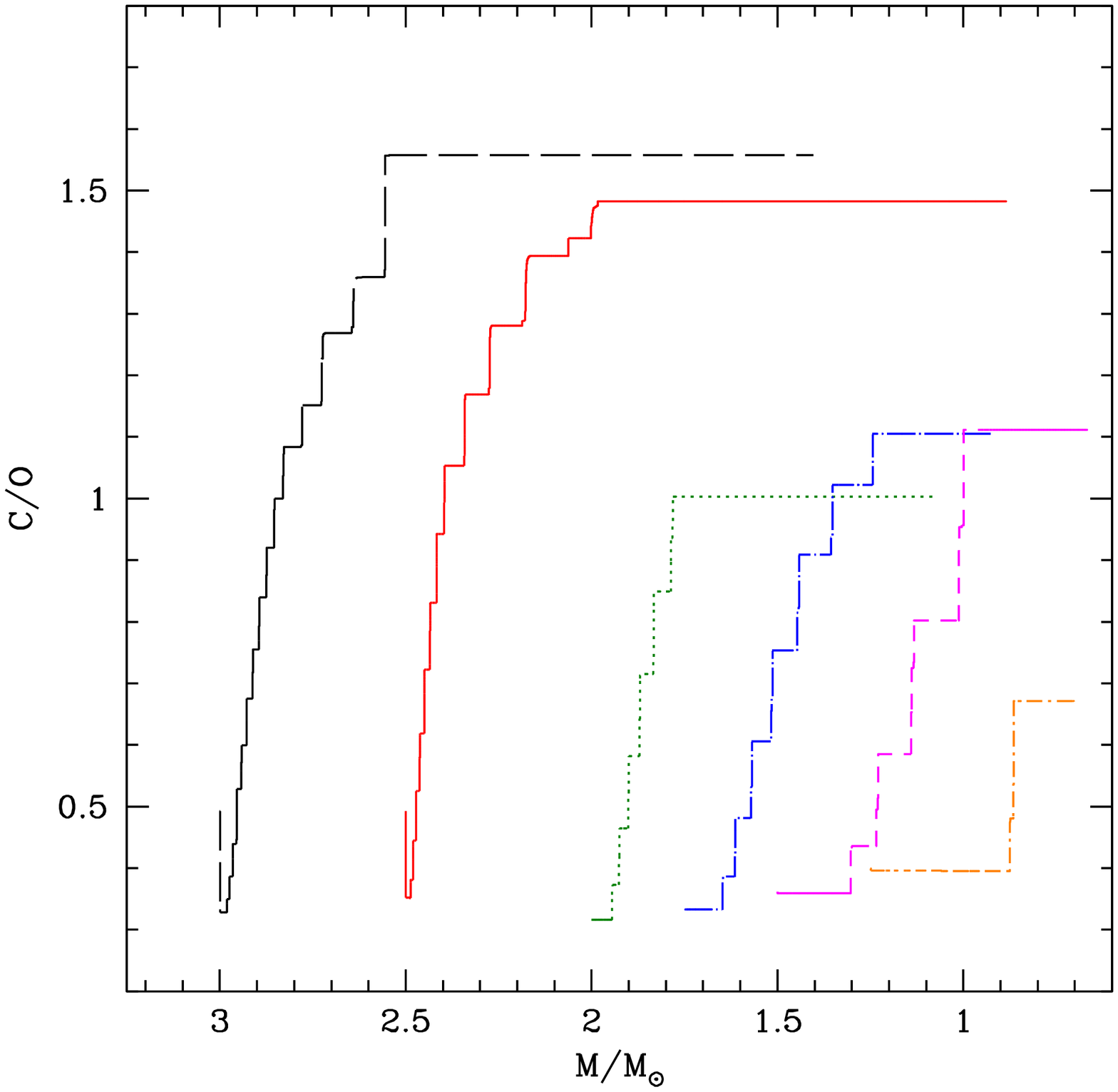}}
\end{minipage}
%\vskip-50pt
\caption{The main physical and chemical properties of low-mass ($M_{init} \leq 3~\Msun$) AGB stars
are shown as a function of decreasing initial mass. Individual panels show the behaviour of luminosity
(top, left), effective temperature (top, right), mass loss rate (bottom, left)
and C/O ratio (bottom, right). The tracks in the panels refer to models of 
initial mass $1.25~M\sun$ (dotted, short-dashed, orange), $1.5~M\sun$ (short-dashed, 
magenta), $1.75~M\sun$ (dotted, long-dashed, blue), $2~M\sun$ (dotted, green), 
$2.5~M\sun$ (solid, red), $3~M\sun$ (long-dashed, black), 
}
\label{fcstar}
\end{figure*}

%%%%%%%%%%%%%%%%%%%%%%%%%%%%%%%%%%%%%%%%%%%%%%%%%%%%%%%%%%%%%%%
\subsection{Low mass AGB stars}
\label{lowmass}

The stars with initial mass below $3.5\Msun$ do not experience any HBB, thus their
chemical composition is entirely determined by the repeated TDU events that follow
each TP. This is going to affect not only their variation
of the surface chemistry, but also their physics.

The main quantities related to the evolution of low initial mass AGB stars are shown in
Fig.~\ref{fcstar}, where we report the variation of the luminosity, effective
temperature, mass loss rate and the surface C/O ratio during the AGB phase.

The C/O ratio evolution shows that after each TP some carbon is dredged-up to the 
surface increasing the C/O ratio. Only stars with initial mass greater than $1.5\Msun$
reach the C-star stage; lower mass stars, while experiencing some
carbon enrichment, lose the external mantle before C/O exceeds unity.

Reaching the C-star stage has important effects on the evolution of these 
objects. As shown in Fig.~\ref{fcstar}, the external regions of the star undergo a 
considerable expansion after the C/O ratio grows above unity: the effective temperature
drops initially to $\sim 3500$ K and decreases further below $3000$ K while the
surface carbon abundance increases. This behaviour is a consequence of the formation of CN
molecules in C-rich regions, that favours a considerable increase in the opacity and 
 in the mass loss rates. This effect was predicted in a seminal
paper by \citet{marigo02} and confirmed in more recent, detailed explorations by 
\citet{vm09, vm10}.

\begin{figure*}
\begin{minipage}{0.49\textwidth}
%\resizebox{1.\hsize}{!}{\includegraphics{figcoratc.ps}}
\resizebox{1.\hsize}{!}{\includegraphics{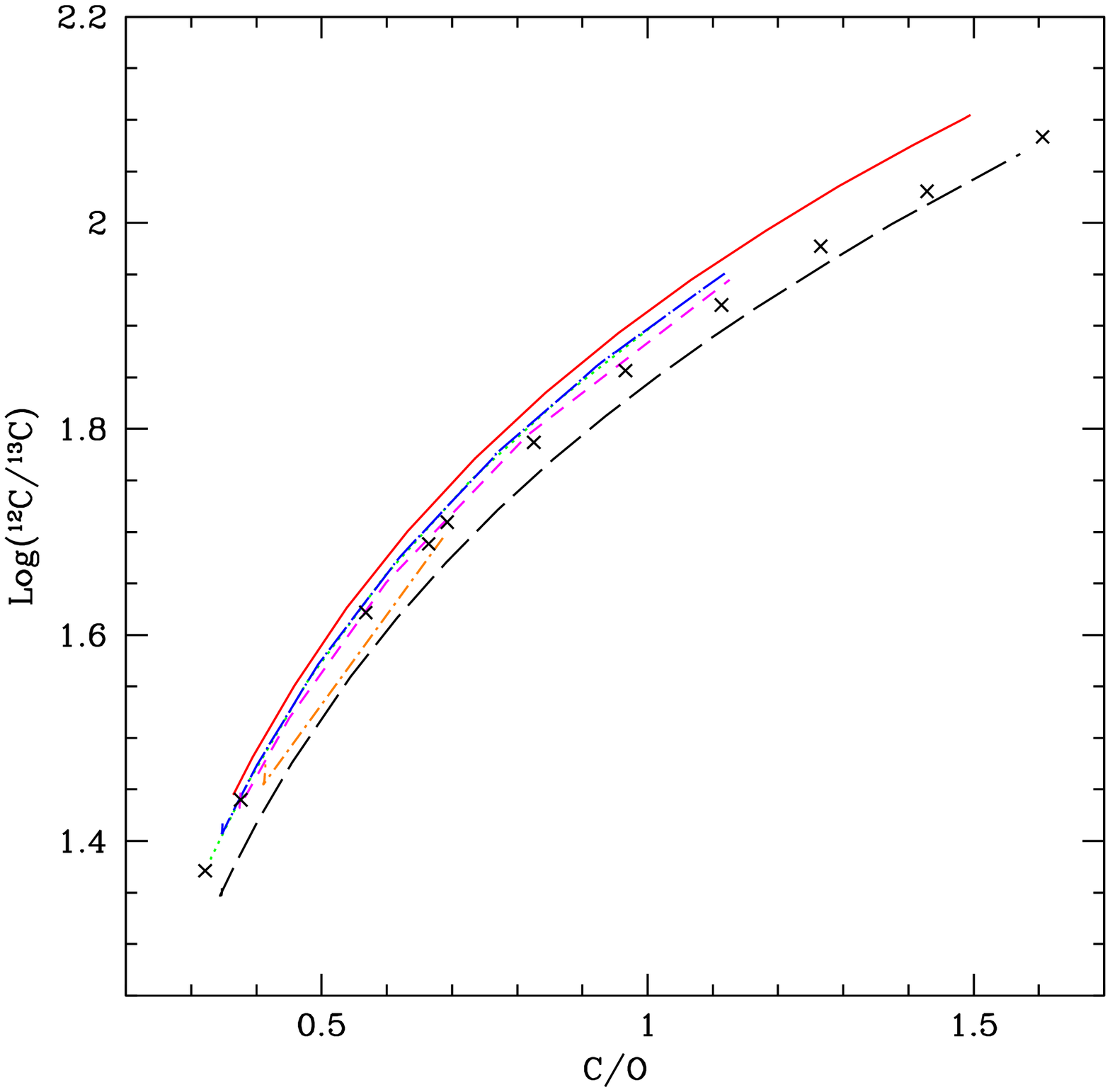}}
\end{minipage}
\begin{minipage}{0.49\textwidth}
%\resizebox{1.\hsize}{!}{\includegraphics{figcolum.ps}}
\resizebox{1.\hsize}{!}{\includegraphics{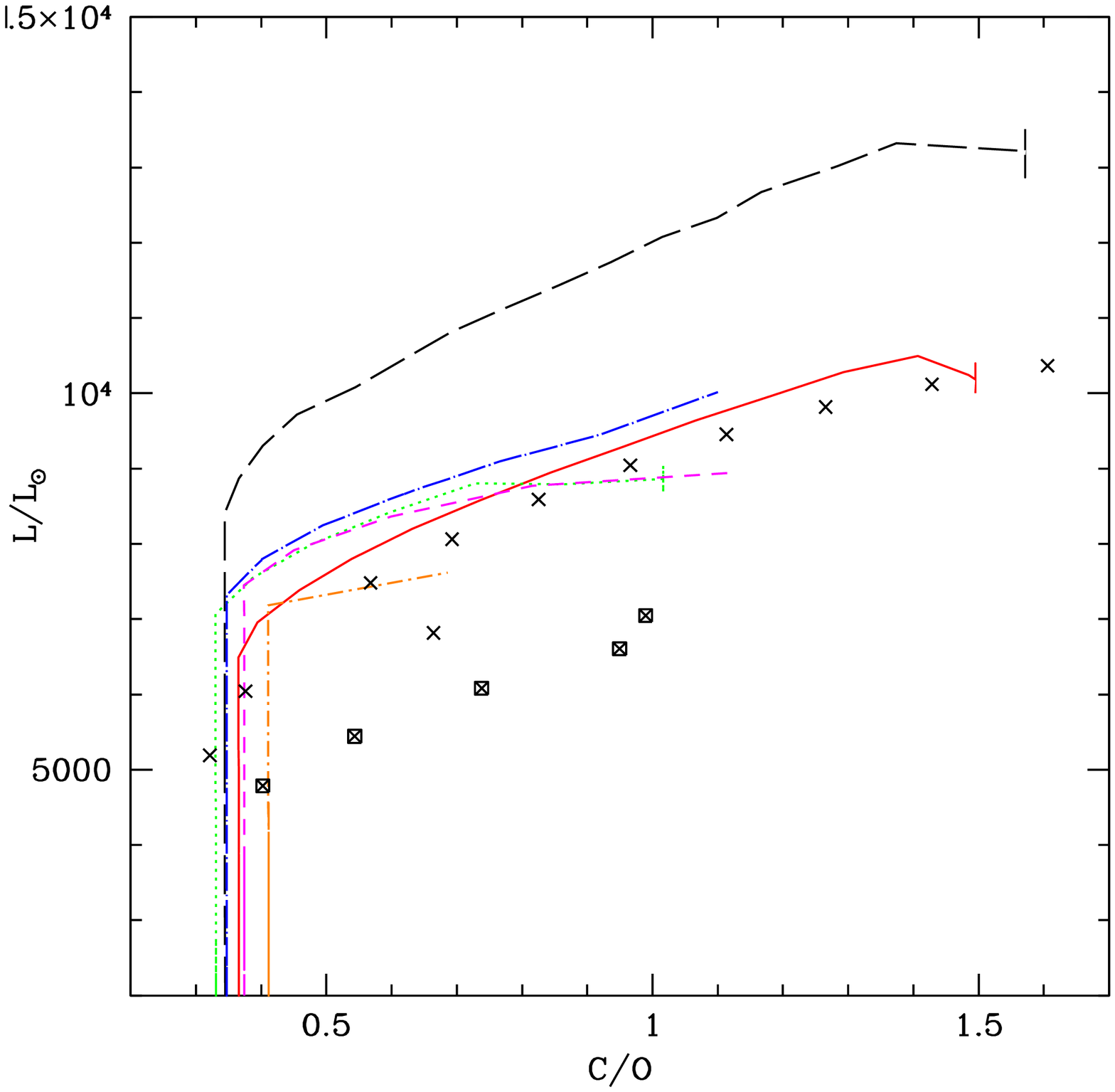}}
\end{minipage}
%\vskip-50pt
\caption{The variation of the surface $^{13}C/^{12}C$ ratio (left panel) and of the 
luminosity (right) for models with mass $M_{init} \leq 3~\Msun$ during the AGB phase.
The two quantities are shown as a function of the surface C/O ratio. The same color coding
of Fig.~\ref{fcstar} was adopted. Crosses and crossed squares refer to C15 models
with initial mass $3~\Msun$ and $1.5\Msun$, respectively.}
\label{fco}
\end{figure*}

As shown in the left bottom panel of  Fig.~\ref{fcstar}, when stars become carbon rich, their mass loss rates 
increase up to $\sim 2\times 10^{-4} \Msun$/yr in the very final phases. 
The increase in the mass loss rate is due to two different effects:  a) the expanded envelope becomes less 
and less gravitationally bound, thus overcoming the gravitational pull is easier
and  b) the lower effective temperatures favour the formation of large quantities
of carbon dust in the wind, which in turn increases the effects of the radiation pressure on the
dust particles in the circumstellar envelope.

Given the above, it is clear why the evolutionary time scales become significantly shorter when stars become C-rich:
the envelope is lost rapidly, only a few (if any) additional TDU events can occur to further increase the surface 
carbon abundance.

The models with mass close to the threshold required to activate HBB, namely
$M_{init} \sim 2.5-3~\Msun$, undergo a higher number of TDU events before their mantle is lost. Consequently,
they are the stars with the largest relative duration of the C-star phase
($\sim 15\%$) and with the highest final C/O ratio ($C/O \sim 1.5$, see Table 1).

%are those experiencing the largest enrichment of carbon, i.e. $C/O \sim 1.5$; this is because they undergo a higher number of TDU events before the envelope is lost. we expect that these stars are those for which the relative duration of the C-star phase, $\sim 15\%$ of the overall AGB evolution, is larger. The duration of the C-rich phase is found to be at most $15\%$ of the total AGB evolution, as reported in Table 1. The largest C/O, found in the models of initial mass $2.5-3 \Msun$, is about 1.5 (see right, bottom panel of Fig.~\ref{fcstar}, where the increase in the rate of mass loss can be deduced by the length of the horizontal part of the tracks in the final phases).

\begin{figure*}
\begin{minipage}{0.49\textwidth}
%\resizebox{1.\hsize}{!}{\includegraphics{figyieldcno_a.ps}}
\resizebox{1.\hsize}{!}{\includegraphics{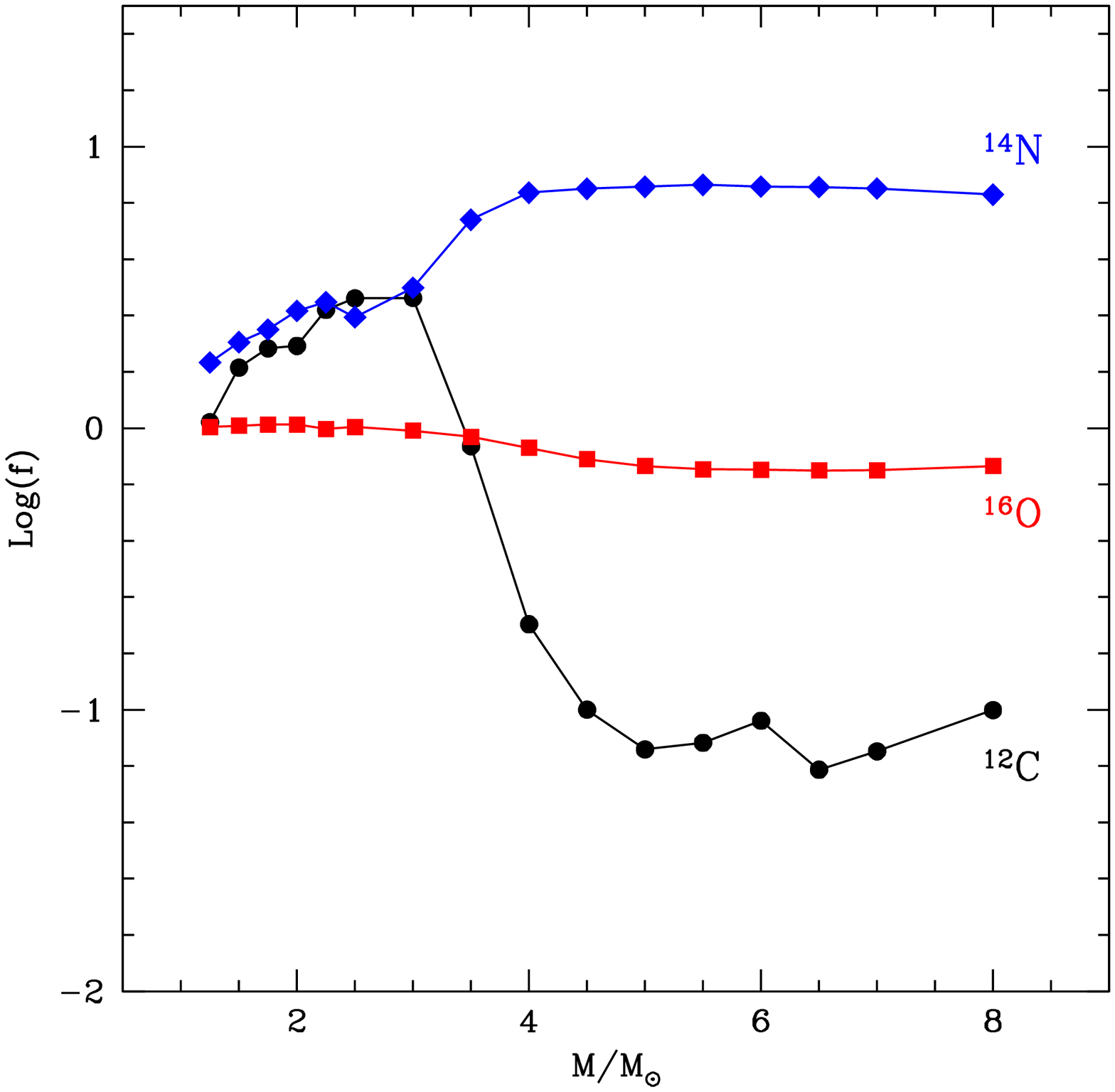}}
\end{minipage}
\begin{minipage}{0.49\textwidth}
%\resizebox{1.\hsize}{!}{\includegraphics{figyieldcno_b.ps}}
\resizebox{1.\hsize}{!}{\includegraphics{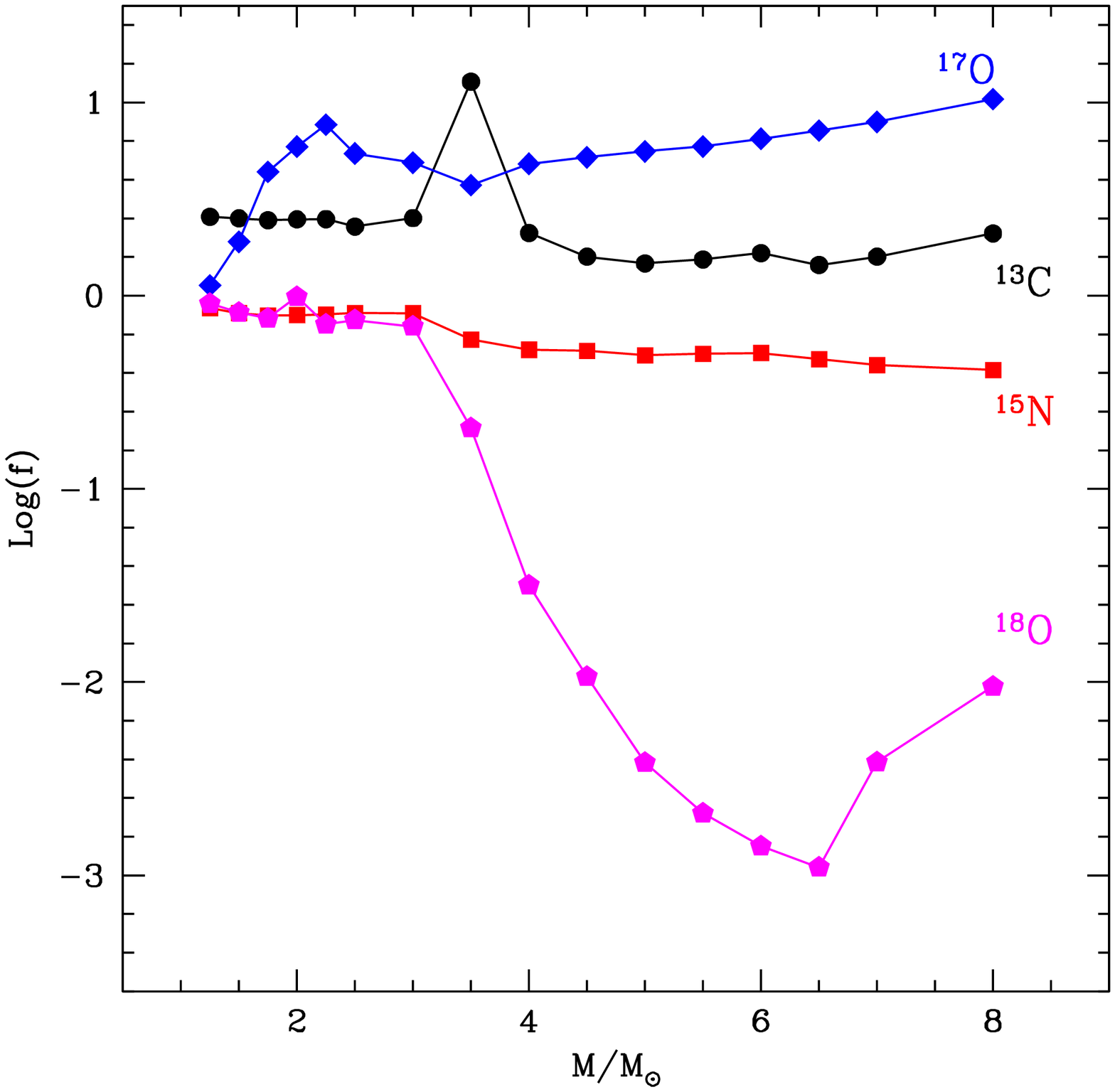}}
\end{minipage}
%\vskip-50pt
\caption{The production factor (see text for definition) of
the CNO isotopes in solar metallicity models. In the left panel we show the most abundant
species, namely  $^{12}C$ (black points), $^{14}N$ (blue diamonds) and $^{16}O$ (red
squares). The right panel refers to $^{13}C$ (black points), $^{15}N$ (red squares),
$^{17}O$ (blue diamonds) and $^{18}O$ (magenta pentagons).
}
\label{fycno}
\end{figure*}

Fig.~\ref{fco} shows the surface $^{12}C/^{13}C$ ratio and the luminosity of the models 
becoming carbon stars during the AGB evolution, as the surface C/O, shown on the 
abscissa, increases. These results show that carbon stars are expected to evolve at
luminosities $8000L_{\odot} < L < 12000L_{\odot}$. Furthermore, the surface $^{12}C/^{13}C$
ratio is expected to be above 50.\\
From the above arguments we understand that the evolution of the C-rich AGB stars is mainly driven by the surface C/O ratio, the latter quantity affecting directly the rate at which mass loss occurs, thus the time scale of this phase.\\
 This is a welcome result for what concerns the robustness of the present findings.
 The increase in the C/O ratio depends on the treatment of
convective borders during each TP, particularly of the assumed extra-mixing from the base
of the envelope and the boundaries of the pulse driven convective shell; however, although
a deeper overshoot would favour larger carbon abundances, this would be counterbalanced
by the increase in the rate of mass loss, which would lead to an earlier consumption of
the stellar mantle, thus reducing the number of additional TDU events.\\
On the statistical side, it is much more likely to detect a star when it is oxygen-rich 
than during the C-star phase. On the other hand, as will be discussed in next section,
the latter is much more relevant for the gas and dust pollution determined by these objects, 
because, as shown in the bottom right panel of Fig.~\ref{fcstar}, it is during this phase that
most of the mass loss occurs.\\

%%%%%%%%%%%%%%%%%%%%%%%%%%%%%%%%%%%%%%%%%%%%%%%%%%
\section{Gas pollution}
The pollution from AGB stars is determined by the relative importance of HBB and TDU
in modifying the surface chemical composition. 

When HBB prevails, N-rich and C-poor yields are expected independently from the HBB strength.
However at high temperatures ($T_{bce} > 80$ MK), when the full CNO cycle and the Ne-Na and Mg-Al chains
can occur, a modification of the mass fraction of elements heavier than oxygen is also expected.
On the other hand, when TDU prevails C-rich yields are expected with minor contribution from O and N.
Table 2 shows the net yields of the various chemical species.
The production factor of the CNO elements, defined as the ratio between the average mass
fraction of a given element in the ejecta and its initial quantity, are shown in 
Fig.~\ref{fycno}. The left panels refers to $^{12}C$, $^{14}N$ and $^{16}O$, whereas
on the right we show the less abundant isotopes.\\
In the low-mass regime ($M \leq 3\Msun$) we find production of $^{12}C$ and $^{14}N$. 
The production factor of both elements increases with the initial mass, up to a maximum 
of $\sim 3$ for $M = 3\Msun$. For what concerns carbon, as discussed in section 3.2, the reason is that higher mass 
models are exposed to more TDU events and experience a larger
enrichment of carbon in the external regions (see bottom, right panel of Fig.~\ref{fcstar}).
%Although the behaviour of carbon and nitrogen is similar, the mechanisms leading to the enhancement are different: while the increase in carbon is due to repeated TDU events, the rise of the surface nitrogen  determined by the first dredge-up episode. 
The null production of carbon found in the
$1.25 \Msun$ model stems from the balance between the first dredge-up, after which the
surface carbon diminishes, and the following TDU's, which increase $^{12}C$ in the
external regions. The first dredge-up is also responsible for the production of 
$^{13}C$ and $^{17}O$ in low-mass AGB stars (see right panel of Fig.~\ref{fycno}): in 
the first case the production factor is $\sim 2.5$, fairly independent of $M_{init}$, 
whereas for the latter isotope it reaches $\sim 10$ in the $2\Msun$ model. $^{15}N$ and 
$^{18}O$ are practically untouched in these stars.

In the high-mass domain the effects of HBB take over, changing the above picture
substantially. Concerning the elements involved in CNO cycling, the results shown in Fig.8 can be
understood based on the discussion in section 3.1.1.
$^{12}C$ is found to be 
10 times smaller in the ejecta, compared to the initial chemical composition. $^{16}O$ is
also affected by HBB, with a maximum depletion of $\sim 30\%$. The CNO
nucleosynthesis has the effect of synthesising $^{14}N$, which results to be increased
by a factor $\sim 8$. 
%While the results regarding $^{12}C$ and $^{14}N$ are in fair agreement with our findings on more metal poor stars, here $^{16}O$ is only modestly affected by HBB, given the lower temperatures at the base of the convective zone of solar metallicity, AGB stars (see top, right panel of Fig.~\ref{fagb}).
The activation of the HBB nucleosynthesis has also the
effects of producing $^{13}C$ and $^{17}O$ via proton capture by $^{12}C$ and
$^{16}O$ nuclei. Note that the significant production of $^{17}O$ (up to a factor $\sim 10$
in the most massive models) is not in contrast with the soft depletion of $^{16}O$,
given the disparity between the initial abundances of the two elements, which renders
a small percentage destruction of $^{16}O$ sufficient to produce $^{17}O$. 
%An additional effect of HBB is the activation of proton capture reactions by $^{18}O$, which is seen to be severely depleted in the ejecta of these stars.
$^{18}O$ is severely depleted in the ejecta of these stars,
being 2-3 orders of magnitude smaller than the initial quantity

Turning to Ne-Na elements, the corresponding production factors are shown in
the left panel of Fig.~\ref{fynena}. We find that $^{22}Ne$ increases in low-mass stars
($M_{init} \leq 3~\Msun$), as a consequence of TDU, which brings to the surface matter 
enriched in $^{22}Ne$; similarly to carbon, the production factor of $^{22}Ne$ is correlated
to $M_{init}$, ranging from $f(^{22}Ne) = 2$ for $M_{init} = 1.5~\Msun$ to $f(^{22}Ne) = 6$
for $M_{init} = 3~\Msun$. Conversely, sodium is only scarcely touched, in this mass interval.

Like in the case of the CNO elements, the transition to the high-mass domain marks an abrupt
change in the surface abundances of the Ne-Na elements, in conjunction with the 
shift from TDU- to HBB-dominated chemistry. 
%$^{22}Ne$ is severely destroyed by HBB, which explains why
In agreement with the discussion in section 3.1.2, we find for $^{22}Ne$  that
 the ejecta of $M_{init} > 3.5\Msun$ exhibit depletion factors ranging from 
3 to 10. Note that the trend of $f(^{22}Ne)$ with mass is not monotonic, the most
$^{22}Ne$-poor ejecta being produced by $M_{init} \sim 6~\Msun$ models, despite the
stronger HBB experienced by their higher mass counterparts. This is motivated by the very
large mass loss rates suffered by $6-8~\Msun$ stars, which render the loss of the envelope
fast enough to compete with $^{22}Ne$ destruction.

The reduction of the surface $^{22}Ne$ favours the production of sodium,
which is increased by a factor $3-5$ in the gas expelled from these stars.
%Compared to the models of lower metallicity, for the solar chemistry we find higher  production factors for sodium, because the temperatures at the base of the envelope are not sufficiently large to destroy the sodium produced by $^{22}Ne$ burning (see Fig.~\ref{fsodio}).
The most abundant isotope of neon, namely $^{20}Ne$, is found to remain practically
unchanged in all cases.

Finally, we examine the Mg-Al elements, shown in the right panel of Fig.~\ref{fynena}. 
In the low-mass domain the surface mass fraction of these elements are only modestly
changed, thus the corresponding production factor are close to unity. In models
of higher mass proton-capture nucleosynthesis occurs, but for the solar metallicity 
the HBB temperatures are not sufficiently large to allow significant depletion of
$^{24}Mg$, which is the starting reaction of the whole cycle: as shown in the right
panel of Fig.~\ref{fynena}, the $^{24}Mg$ in the ejecta is barely depleted by more
than $\sim 25\%$, even in the models of highest mass. This partial nucleosynthesis
is however sufficient to produce $^{25}Mg$, which is found to be increased by a factor
$\sim 4$ in the most massive models.

%%%%%%%%%%%%%%%%%%%%%%%%%%%%%%%%%%%%%%%%%%%%%%%%%%%%%%%%%
\section{The robustness of the present generation of AGB models}
The results from AGB evolution modelling are sensitive to the treatment of some
physical mechanisms still poorly known from first principles, primarily convection 
and mass loss. Additional uncertainties come from the nuclear reactions cross-sections,
though this is going to affect only the details of the chemical composition of the
ejecta, because the nuclear rates of the reactions giving the most relevant contribution
to the overall energy release are fairly well known \citep{herwig05, karakas14b}.

A reliable indicator of the predictive power of the present findings can be obtained
by comparing with solar metallicity, AGB models found in the literature. 
On this purpose, in Fig.~\ref{fagb},
reporting the main physical properties of the models presented here, we also show
the results from \citet{cristallo15}, \citet{karakas16}, \citet{doherty14}. In the
following, we will refer to the four sets of models, respectively, as ATON, C15,
K16 and D14.

In the low-mass regime, the main difference between ATON and C15 and K16 results is 
that the ATON core masses at the beginning of the AGB phase are slightly 
smaller, thus the ATON models evolve at lower luminosities and the AGB phase is longer. 
The largest difference is found for the $2\Msun$ model, which in the ATON case experiences
a maximum luminosity $0.08$ dex smaller than C15, and the AGB evolution is two times
longer (4Myr vs. 2Myr). K16 models exhibit an intermediate behaviour in this range of mass.

\begin{figure*}
\begin{minipage}{0.49\textwidth}
%\resizebox{1.\hsize}{!}{\includegraphics{figyieldnena.ps}}
\resizebox{1.\hsize}{!}{\includegraphics{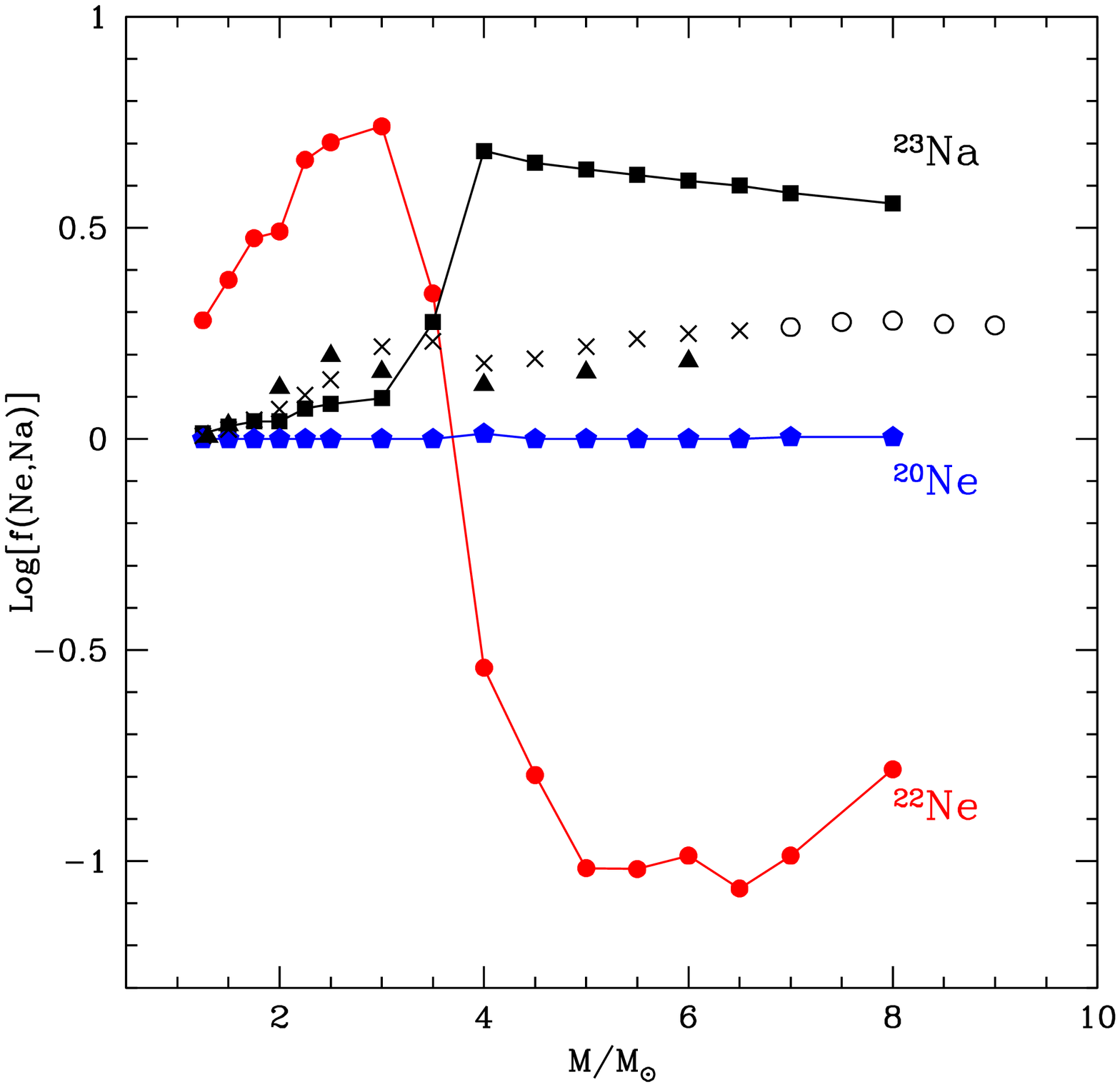}}
\end{minipage}
\begin{minipage}{0.49\textwidth}
\resizebox{1.\hsize}{!}{\includegraphics{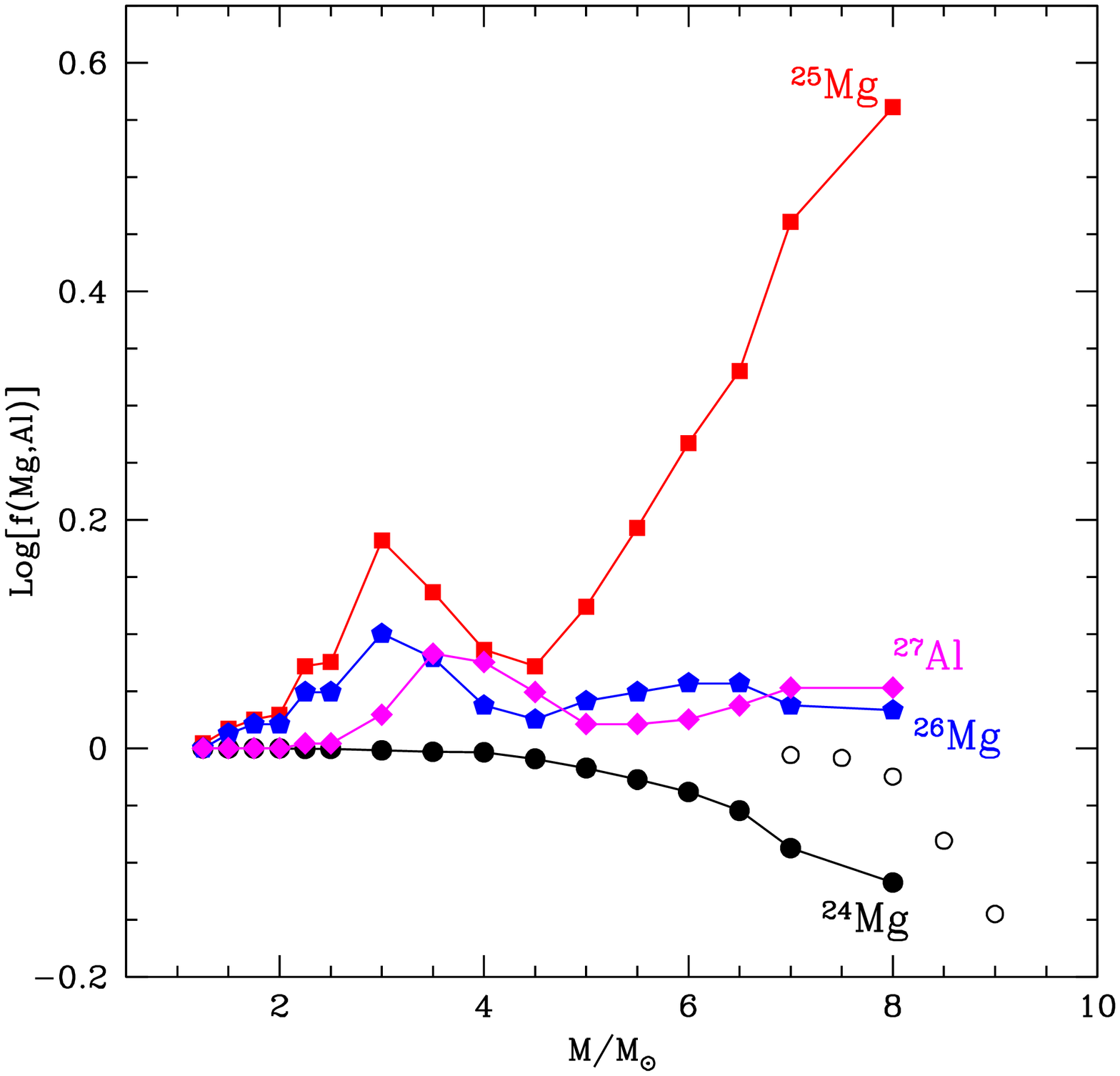}}
%\resizebox{1.\hsize}{!}{\includegraphics{figyieldmgal.ps}}
\end{minipage}
%\vskip-50pt
\caption{The production factor of the elements involved in the Ne-Na and Mg-Al nucleosynthesis
for the models presented here. Left: $^{20}Ne$, $^{22}Ne$ and $^{23}Na$ are indicated, respectively,
with blue pentagons, red points and black squares; the sodium production factor by K16
(crosses), c15 (triangles) and D14 (circles) are also indicated. Right: the production
factor of $^{24}Mg$ (black points), $^{25}Mg$ (red squares), $^{26}Mg$ (blue pentagons)
and $^{27}Al$ (magenta diamonds); the results for $^{24}Mg$ by D14 are indicated with 
open circles.}
\label{fynena}
\end{figure*}

The most relevant differences are found in the high-mass domain, where HBB effects take over.
In the comparison among the highest temperatures reached at the base of the convective
envelope, ATON models in the range $4-6\Msun$ attain values of the order of $80-90$ MK,
whereas in the C15 case we find $10MK < T_{bce} < 20MK$. The K16 models exhibit 
temperatures closer to, though smaller than ATON, covering the range 
$30MK < T_{bce} < 80MK$ in the same interval of mass.
Such a dramatic difference has an immediate effect on the luminosity, which for the 
ATON models, in the same range of mass, is $30000 < L/L_{\odot} < 60000$, whereas in 
the C15 and K16 cases it is, respectively, $20000 < L/L_{\odot} < 30000$ and
$20000 < L/L_{\odot} < 40000$. Because the core masses at the beginning of the AGB
phase are very similar in the three cases (see Fig.1), the differences outlined
above must originate from the different description of the convective instability, particularly
for what concerns the efficiency of convection in the innermost regions of the
envelope. The ATON models are based on the FST treatment \citep{cm91}, whereas the C15 and 
K16 computations used the mixing length theory (MLT) recipe. These results confirm the abalysis by \citet{vd05a}, 
who discussed the outstanding impact of convection modelling on the
efficiency of HBB experienced by AGB stars.

In the analysis of the 
behaviour of the core masses, we note that the ATON 
models present the greatest variation ($\delta M_C \sim 0.05 \Msun$) during the whole AGB 
phase, compared to C15 and K16, for which we have $\delta M_C < 0.02 \Msun$: this is
due to the deeper penetration of the convective envelope in the phases following
each TP in the C15 and K16 cases, which slows the growth of the core during the AGB evolution.

We now focus on the evolution properties of those stars that develop a core made up of
oxygen and neon, i.e. those of initial mass above $6.5~\Msun$. In this case we compare
the ATON models with D14 and with the $8~\Msun$ model by K16\footnote{The ATON and K16
models of, respectively, $7~\Msun$ and $8~\Msun$, produce indeed an hybrid O-Ne core: 
they undergo an off-centre ignition of carbon, but the temperatures are not sufficient
for the convective flame that develops to reach the centre of the star.}. 
The same initial mass does not
correspond to the same core mass during the early AGB phases, because in the ATON case
a larger extra-mixing from the border of the convective core during the H-burning
phase was adopted, which results into a higher core mass at the beginning of the AGB
phase. Taking into account this difference, we note that the values of the temperature
at the base of the envelope and of the luminosity are similar in the ATON, D14 and K16 cases,
whereas the D14 and K16 AGB evolutionary times are longer than ATON. The results from this 
comparison finds an explanation in the different modalities with which convection
and mass loss are described. ATON models are based on a more efficient
description of convection (FST, against the MLT treatment used by D14 and K16), which 
favours larger luminosities and HBB strength; however, ATON models also suffer a very 
strong mass loss,
which provokes a fast loss of the external mantle, accompanied by a general cooling of
the whole external regions, which acts against the achievement of very large HBB
temperatures. The longer duration of the AGB phase found in the D14 and K16 models is due
to the smaller mass loss rate adopted compared to ATON. The interested reader may find in 
D14 an exhaustive discussion of the impact of the mass loss description on the duration of
the TP phase of super-AGB stars.

\begin{figure*}
\begin{minipage}{0.33\textwidth}
%\resizebox{1.\hsize}{!}{\includegraphics{figc12conf.ps}}
\resizebox{1.\hsize}{!}{\includegraphics{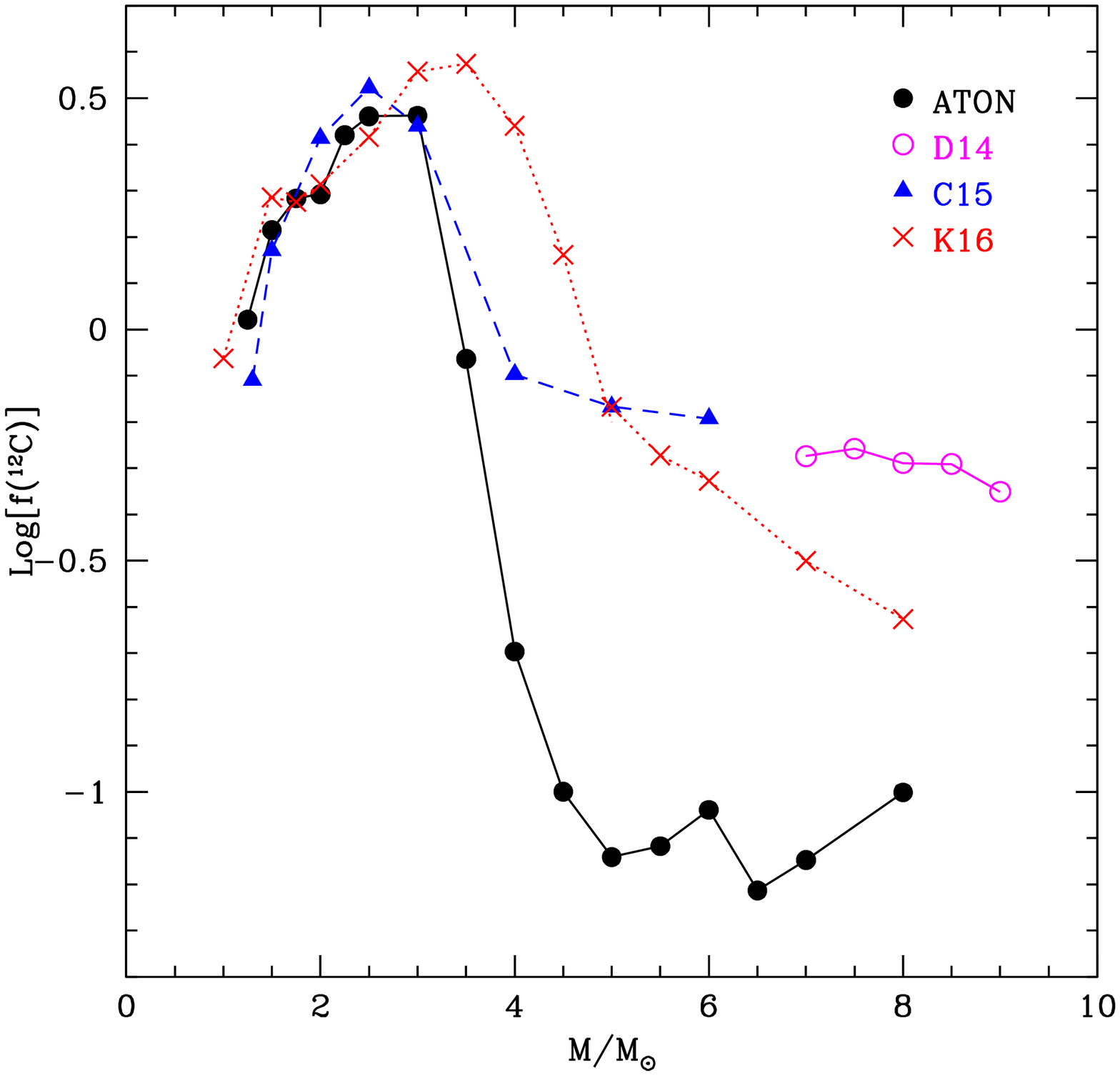}}
\end{minipage}
\begin{minipage}{0.33\textwidth}
%\resizebox{1.\hsize}{!}{\includegraphics{fign14conf.ps}}
\resizebox{1.\hsize}{!}{\includegraphics{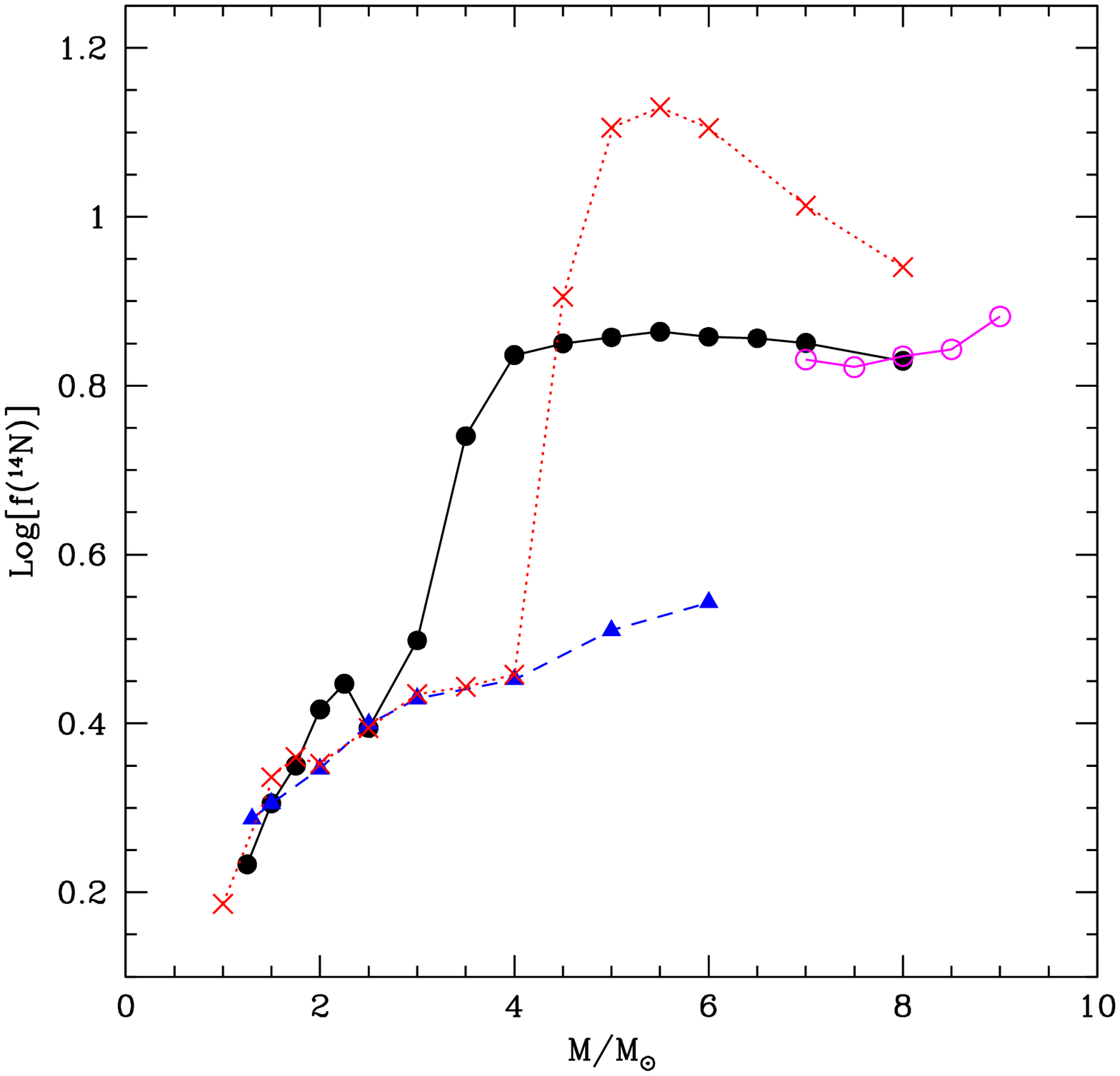}}
\end{minipage}
\begin{minipage}{0.33\textwidth}
%\resizebox{1.\hsize}{!}{\includegraphics{figo16conf.ps}}
\resizebox{1.\hsize}{!}{\includegraphics{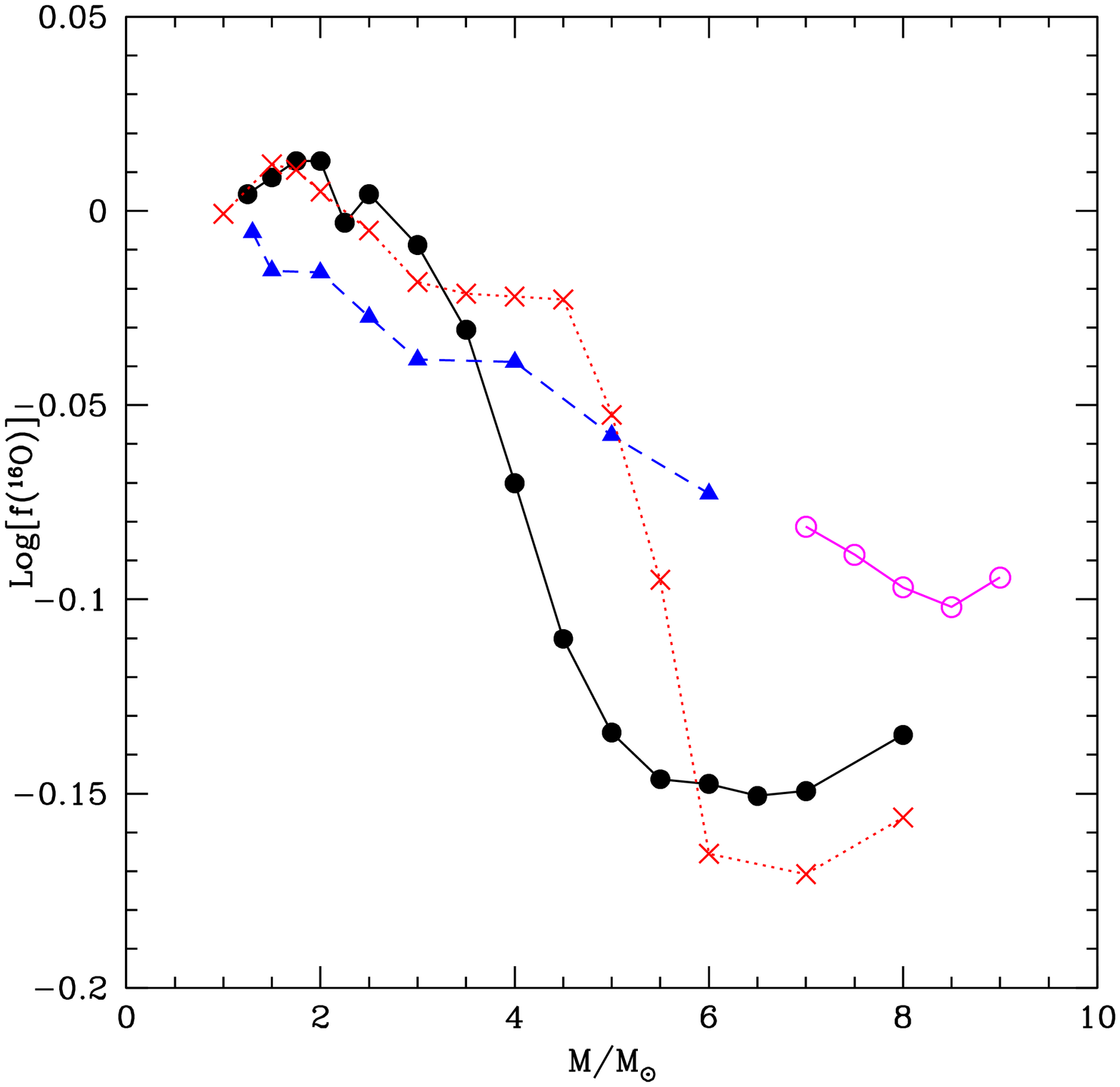}}
\end{minipage}
%\vskip-30pt
\caption{The production factor of $^{12}C$, $^{14}N$ and $^{16}O$ of the models 
presented here (shown as black, full points), compared with results from
\citet{cristallo15} (blue triangles), \citet{karakas16} (red crosses) and
\citet{doherty14} (magenta circles).
}
\label{fcnoconf}
\end{figure*}

The differences discussed above have important effects on the yields expected from 
these stars, which show some differences among the results published by the
different research groups.

In Fig.~\ref{fcnoconf} we compare the
production factors of the most abundant isotopes involved in CNO nucleosynthesis.

In the low-mass domain we find that the results concerning carbon are very similar.
In all cases we find a positive trend of the carbon in the ejecta with the stellar mass,
as higher mass stars experience more TDU events. ATON, C15 and K16 results are also
similar on quantitative grounds, the largest $^{12}C$ enhancement being $\sim 3$, reached
by $\sim 3~\Msun$ models.

The N-production factor of  ATON, C15 and K16 are also similar: $f(N)$ increases
increase with the stellar mass, up to $f(N) \sim 3$ for the $\sim 3~\Msun$ models.

In the same range of mass a few differences are found for what regards oxygen.
In the ATON and K16 cases some oxygen enrichment occurs, whereas no $^{16}O$ production is
found in C15 models.

For what concerns stars of mass above $3\Msun$, the predictions are considerably
different. In the ATON case carbon in the ejecta is severely reduced, almost
by a factor 10. In the C15 and K16 models this reduction is much smaller,
at most by a factor 4 in the $8~\Msun$, K16 model. The results from D14 also predict
reduction factors not higher than 2. 
 
Concerning nitrogen, in the mass range 
$3~\Msun < M <4~\Msun$ the ATON models produce more nitrogen, owing to the effects of
HBB, not found in the C15 and K16 models of the same mass.
In the ATON case, for massive stars, great amount of nitrogen are produced, with production 
factors in the range 6-8. This behaviour is shared by the D14 models. Conversely, 
the N-production factor is significantly smaller in the C15 case, where the
production factor never exceeds $\sim 4$. The largest production of nitrogen is found in 
the $5-6~\Msun$ models by \citet{karakas16}: this is 
due to the combined effects of TDU, which increases the surface carbon, and HBB, which 
converts the dredge-up carbon into nitrogen.

In the large mass domain $^{16}O$ is only modestly reduced in C15 models, whereas in the ATON case the 
depletion factor in $M_{init}\geq 5\Msun$ models is $\sim 70\%$. The comparison between
the ATON and K16 models is more tricky: for $M_{init} < 6~\Msun$ the ATON models predict 
more oxygen-poor ejecta, whereas in the range of mass $6~\Msun < M <8~\Msun$ the oxygen 
depletion is slightly higher in the K16 case. In the D14 models some oxygen depletion is 
found, though limited to $\sim 20\%$.

Turning to sodium, in the large mass domain the results are significantly different,
as can be seen in the left panel of Fig.~\ref{fynena}: in the ATON case a great production 
of sodium is expected, the average Na in the ejecta being increased, with respect to 
the original chemistry, by a factor ranging from 3 to 5. In the K16 and D14 models the 
production factor is below 2, whereas in the C15 case it is slightly smaller. 

In the range of masses experiencing HBB the extent of the Mg-Al nucleosynthesis is
also model-dependent, as shown in the right panel of Fig.~\ref{fynena}. The ATON models
achieve some processing of $^{24}Mg$, which is depleted by at most $\sim 40\%$ in the
most massive case. This is in fair agreement with the results from D14, whereas in the
C15 and K16 models processing of magnesium is negligible.

The differences in the expected chemical enrichment of the interstellar medium can be
understood on the basis of the physical input used by the various research groups 
to calculate the evolutionary sequences. Convection is by far the biggest villain here,
determining most of the differences found. 

In the low-mass domain, the slight increase in the $^{16}O$ content found in the ATON
and K16 models is due to the adoption of some overshoot from the base of the pulse driven 
convective shell, which further enhances the strength of the pulse and, more important,
makes the internal regions of the convective envelope to be mixed with more internal 
zones touched by helium burning, with a higher oxygen content. In the same range of mass
we find that the largest production factor of $^{12}C$ is similar in the ATON, C15 and
K16 cases, indicating that those models experience TDU events of similar depth.

For masses above $3\Msun$, the main reason for the differences among the various models
is the strength of HBB and the description of mass loss. The large HBB temperatures are 
the main actors in the considerable depletion of carbon and  production of nitrogen in 
the ATON models. The K16 models of mass above $5\Msun$ and the D14 models produce ejecta 
with nitrogen enhancement similar
to ATON (see middle panel of Fig.~\ref{fcnoconf}), despite the carbon depletion is
more reduced (left panel of the same figure). This is motivated by some TDU events
active in the latter models, that transport to the external regions some carbon produced
in the helium-burning shell, which is later converted into nitrogen: in summary, while
the nitrogen produced in the ATON models is entirely of secondary origin, part of the
nitrogen synthesised in the K16 and D14 cases has also a primary component.
The best indicator of the efficiency of HBB is the behaviour of oxygen, which is depleted 
in the ejecta of the ATON and in some K16 models, whereas it is only scarcely touched in 
the C15 and D14 cases (see right panel of Fig.~\ref{fcnoconf}). 
Understanding the differences among the ATON and K16 results is not straightforward though.
For $M_{init} < 6~\Msun$ the
ATON models predict more oxygen-poor ejecta, because the K16 models are cooler at the
base of the envelope (see top, right panel of Fig.~\ref{fagb}), thus the latter is exposed 
to a less advanced nucleosynthesis. In the range of mass $6~\Msun < M <8~\Msun$ the oxygen 
depletion is slightly higher in the K16 case, compared to ATON, despite the latter models 
evolve at larger $T_{bce}$'s. The reason for this apparently anomalous behaviour is once
more in the large mass loss rates experienced by the ATON models, which makes the envelope
to be lost before a great depletion of the surface oxygen may have occurred.

The efficiency of HBB is also the main factor determining the extent of the Ne-Na and
Mg-Al nucleosynthesis experienced. The great enhancement of sodium found in the ejecta
of $M\geq 3.5\Msun$ ATON models is originated by the large HBB temperatures reached;
conversely, in the other cases the temperatures required to activate the Ne-Na
nucleosynthesis are barely reached, which determined a much smaller production of
sodium (see left panel of Fig.~\ref{fynena}).

\section{Interpretation of observed Galactic AGB stars}
The discussion of the previous sections outlines how far we are from a full understanding
of the main evolutionary properties of AGB stars. The significant differences found
between the present models and those by K16, C15 and D14 stress the importance of 
comparing the expectations from the models with the observations. 
%On this point of view,  this research will soon benefit of the accurate distance determinations of Galactic AGB  stars, obtained via the Gaia space mission. The knowledge of the distance will allow a robust determination of the luminosity, which, as discussed in the previous sections, proves crucial to the identification of the stars observed. 
As a first step towards this
direction, we compare the most recent estimates of the CNO elemental and
isotopic abundances in Galactic (solar metallicity) AGB stars with the ATON
models presented here.

\subsection{Extreme O-rich, AGB stars observed by Herschel}
\label{just}

\begin{figure*}
\begin{minipage}{0.49\textwidth}
%\resizebox{1.\hsize}{!}{\includegraphics{figratc.ps}}
\resizebox{1.\hsize}{!}{\includegraphics{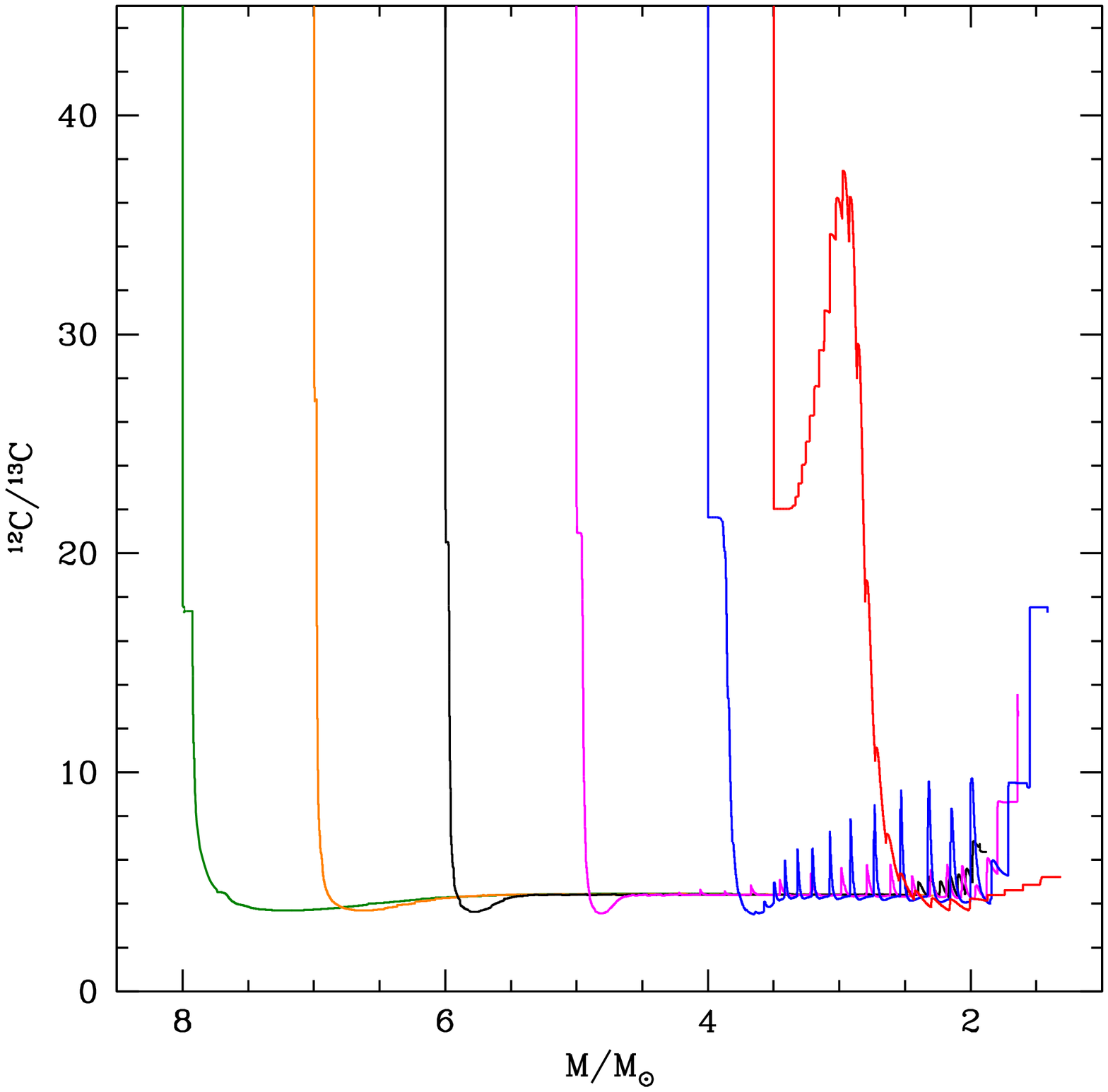}}
\end{minipage}
\begin{minipage}{0.49\textwidth}
%\resizebox{1.\hsize}{!}{\includegraphics{figo17o18.ps}}
%\resizebox{1.\hsize}{!}{\includegraphics{figo17o18.ps}}
\resizebox{1.\hsize}{!}{\includegraphics{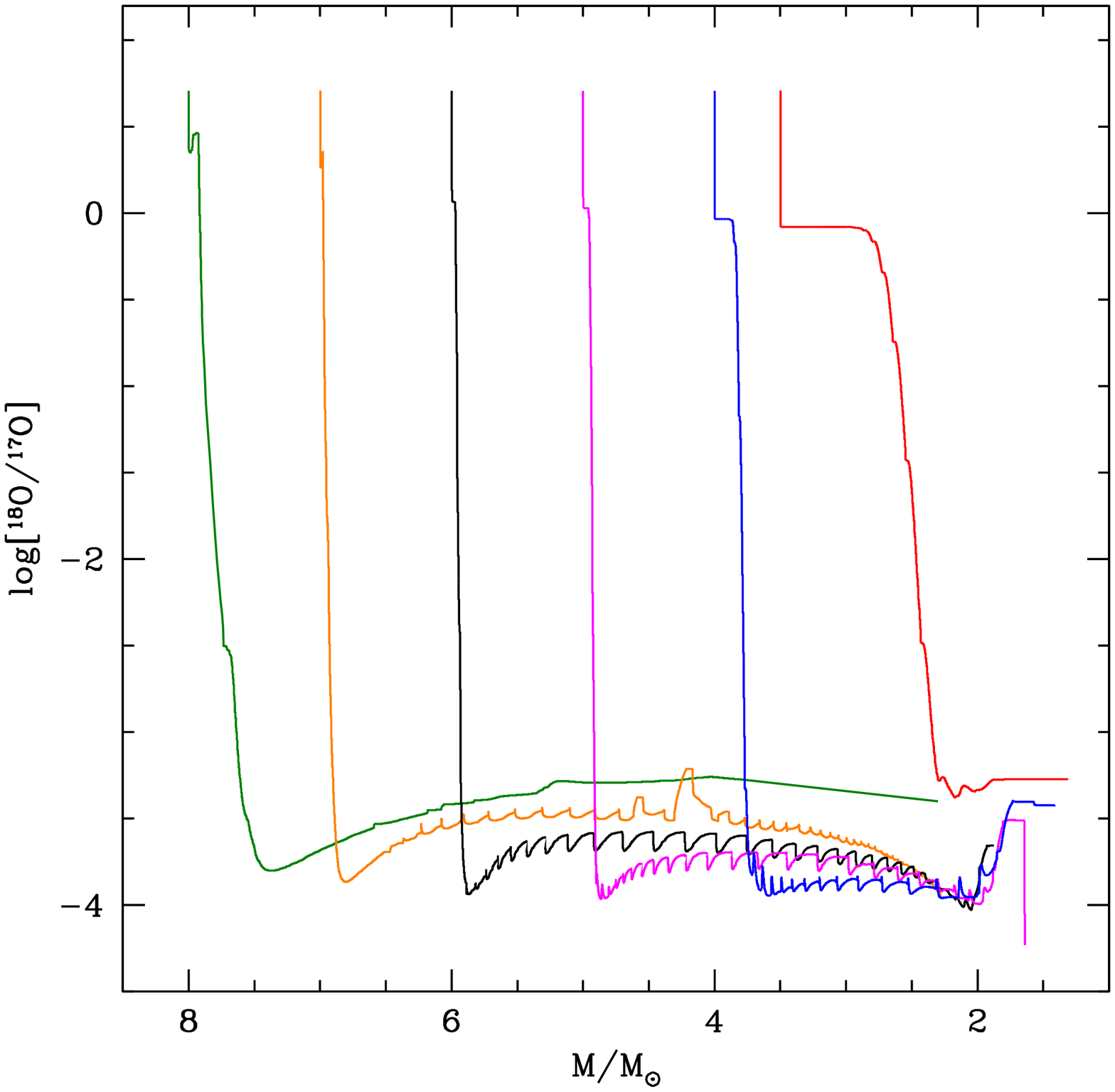}}
\end{minipage}
%\vskip-50pt
\caption{The evolution of the surface $^{12}C/^{13}C$ (left panel) and
$^{18}O/^{17}O$ ratios, in the same models shown in Fig.~\ref{flum}. The same colour 
coding is adopted.
}
\label{fratco}
\end{figure*}

\citet{sample1} published Herschel Space Observatory (Herschel hereafter)
observations of five visually obscured OH/IR stars\footnote{These stars
are obscured in the optical range (e.g. Garcia-Hernandez et al. 2007)
and they are expected to be the more massive AGB stars,
experiencing extreme mass-loss rates.}, using CO as a tracer of
the thermodynamical structure of the circumstellar envelope. The combination with ground data
allowed the determination of the dynamical and dust properties of the wind, and
the derivation of the oxygen and carbon isotopic ratios. \\
To allow a clearer interpretation 
of the chemical composition of the stars in this sample we show in Fig.~\ref{fratco}
the evolution of the surface $^{12}C/^{13}C$ (left panel) and of  $^{18}O/^{17}O$  (right)the same models shown in Figg.~\ref{fagb} and \ref{fcno}. 
In all cases we see a significant reduction of the surface $^{12}C/^{13}C$ as soon as
HBB begins, owing to the destruction of $^{12}C$ and the synthesis of $^{13}C$;
eventually, the equilibrium value, $^{12}C/^{13}C \sim 4$, is reached. The activation
of HBB also determines the destruction of the surface $^{18}O$ and the synthesis of
$^{17}O$: the surface $^{18}O/^{17}O$ is dramatically reduced compared to the initial value, $^{18}O/^{17}O=5$. In the models of initial mass $M_{init} \sim 4~\Msun$ the surface $^{12}C/^{13}C$ raises again in the final evolutionary phases, after HBB was switched off: under the effects of a few late TDU events, carbon ratios $^{12}C/^{13}C \sim 20$ are expected.

The surface chemistry of OH 127.8+0.0 and OH 30.1-0.7 shows the clear imprinting of 
HBB, with $^{12}C/^{13}C \sim 3-5$ and $^{18}O$ below the detectability threshold. 
As shown in Fig.~\ref{fratco}, this is a common feature of all the models experiencing HBB. 
The ignition of $^{12}C$ burning (with consequent synthesis of $^{13}C$) and the 
depletion of $^{18}O$ are active since the early AGB phases; as shown in 
Fig.~\ref{fagb}, these temperatures at the base of the convective envelope are reached 
in all the models experiencing HBB. The surface chemistry observed in OH 127.8+0.0 and 
OH 30.1-0.7 is a common feature of all the models of initial mass 
$M_{init} \geq 3.5\Msun$, thus not allowing us discriminating among the possible 
progenitors. The upper limits for the $^{18}O/^{17}O$ ($\leq$0.1) given by Justtanont et al. (2013) also support our interpretation of these stars being massive HBB stars. 
Our interpretation agrees with the Justtanont et al. (2013) conclusion of
these stars being HBB AGB stars; the difference in the progenitor mass range 
(Justtanont et al. 2013 assume masses above $\sim 5\Msun$) is just because the minimum 
mass to activate HBB is model dependent; e.g., it is $\sim 3.5~\Msun$ in
the ATON models, while it is $\sim 4.5~\Msun$ in the D14-like models used by
Justtanont et al. (2013). We note, however, that these stars might be in a very
advanced evolutionary stage, thus implying that their current mass could in principle
be significantly smaller (down to $\sim 1~\Msun$) than the initial mass. \\
Based on ISO spectra and IRAS photometry, the SED of these two
stars shows up the silicate absorption feature at $9.7\mu$m, in agreement with the 
hypothesis that they are undergoing HBB: indeed \citet{flavia14} showed that large dust
formation occurs in $M_{init} > 3~\Msun$ stars during the HBB phase. The results
by \citet{flavia14} were based on models with sub-solar chemical composition; while
the dust production by the present models will be addressed in a forthcoming paper,
we may anticipate that the conclusions by \citet{flavia14} can be safely extended to the
present case, because the larger availability of silicon in the surface regions will
further increase dust production in solar metallicity stars. We conclude that
OH 127.8+0.0 and OH 30.1-0.7 are evolving through the AGB phases during which HBB is
strongest, when dust production is large and the stars lose mass at high rates.\\
A significant support towards the identification of the
precursors of OH 127.8+0.0 and OH 30.1-0.7 could be obtained by the knowledge of their
distance, which would allow the determination of their luminosity. This is because, while
undistinguishable on the basis of the surface isotopic ratios of carbon and oxygen, the
stars of the various mass evolve at different luminosities during the AGB phase. This
is clearly shown in Fig.~\ref{flum}, where the range of the luminosities of the various
tracks is seen to vary substantially with the initial mass of the star: a luminosity
$L \sim 2\times 10^4 L_{\odot}$ would point in favour of the progeny of $\sim 3.5\Msun$
star, whereas a higher mass progenitor, $\sim 8\Msun$, would require much higher
luminosities, of the order of $L \sim 10^5 L_{\odot}$. It goes without saying that we 
mentioned only the two extreme cases, neglecting a number of intermediate situations.

The surface chemical composition of AFGL 5379 and OH26.5+0.6 indicates depletion of 
$^{18}O$, as confirmed by the non detection of the $H_2^{18}O$ line in the spectra. 
Unlike the two previous stars, the isotopic carbon ratio, $^{12}C/^{13}C \sim 15-20$, 
is significantly higher than the equilibrium value. \\
A possible interpretation of these data is that AFGL 5379 and OH26.5+0.6 descend from
$3.5-4~\Msun$ progenitors and are in the phases following the ignition
of HBB, when carbon burning started, but there was no time to reach the equilibrium 
value. In this case we expect that the current mass of the stars are close to the 
initial mass and that the stars are actually lithium-rich\footnote{This information is 
however of little help because, as we noted above, these sources are completely obscured 
in the optical \citep{garcia07}, leaving no chances of any reliable lithium measurement.}. 
We believe this possibility unlikely,
for the following reasons: a) at the ignition of HBB these stars would evolve at
effective temperatures $T_{eff} \sim 3000$ K, significantly higher than the temperatures
deduced by \citet{sample1}, which are slightly above 2000 K; b) during the same phase, we find that
these stars have radii of the order of $\sim 500R_{\odot}$, $\sim 40 \%$ smaller than
found by \citet{sample1}; c) the mass loss rates in the initial HBB phases are at
most a few $10^{-6} \dot M/yr$, whereas AFGL 5379 and OH26.5+0.6 are currently loosing
mass with rates much higher than $10^{-5} \dot M/yr$.\\
Our favourite interpretation is that AFGL 5379 and 
OH26.5+0.6 are the progeny of $\sim 4~\Msun$ stars, and are currently evolving during 
the final AGB phases. The observed $^{12}C/^{13}C$ is larger than the equilibrium
value, because HBB is switched off when the mass of the envelope drops below
$\sim 1~\Msun$ and a few TDU events are sufficient to increase the surface $^{12}C$, 
thus lifting the $^{12}C/^{13}C$ ratio (see left panel of Fig.11). The effective temperatures during
the late AGB phases are $T_{eff} \sim 2500$ K, in better agreement with those
indicated by the authors, i.e. $T_{eff} \sim 2200$ K. An additional point in favour of
this hypothesis is that the radius of the star is expected to be $\sim 800R_{\odot}$,
very close to the values proposed by \citet{sample1}. A last argument supporting this
conclusion is that the SED of these stars show up a deep silicate feature, suggesting
the presence of significant quantities of dust, as expected based on the cool temperatures
of the models, favouring dust formation. Interestingly, if this hypothesis proves correct, 
it is possible to constrain the current mass and the luminosity of AFGL 5379 and 
OH26.5+0.6: in the final AGB phases of $\sim 4~\Msun$ stars the mass is reduced to 
$1-2\Msun$ and the luminosity is $L=2-3\times 10^4 L_{\odot}$.

This could be confirmed by an accurate determination of the distances which is not yet available, at the moment, for this type of stars.

Among the stars observed by \citet{sample1} WXPsc is the least obscured and is still visible in the optical. The carbon ratio for this star is $^{12}C/^{13}C = 10 \pm 4$;  this is not highly significant, as it ranges from the values typical of CNO equilibria to those of incomplete CN burning. 
The information on the  $^{18}O/^{17}O $ ratio is hard to interpret: Justtanont et al. (2012)  give $^{18}O/^{17}O \sim 1.5$, at odds  with the 
results derived by Justtannont et al. (2015) in a larger sample of extreme OH/IR stars, where they found upper limit for the oxygen isotope ratio of the order of  0.1. An additional information on this star is that the optical spectrum displays a strong Rb line at 7800A (Garcia-Hernandez 2016, private communication), which suggests that it has  already experienced some TPs and TDU episodes, and it is Rb-rich.
Confirmation of the $^{18}O/^{17}O $ given by Justtanont et al. (2012) would rule out any contamination from HBB; in this case
the most likely possibility is that WXPsc descends from a progenitor of mass just above the threshold required to activate HBB
($M_{init} \sim 3.5~\Msun$) and has already experienced some TDU events, whereas HBB has not yet started.

This interpretation has some problems though, mainly related to the degree of obscuration of the star ($\tau_{10}=3$, according to
Ramstedt \& Oloffsson, 2013), as witnessed by the silicate feature, which is about to be converted into absorption, owing to the
increasing thickness of the circumstellar shell: this evidence would rather indicate that WXPsc is evolving through the final AGB 
phases and is surrounded by great quantities of silicate dust. If this understanding is correct, the surface chemistry of the star
should display evidences of HBB, which seems in contrast with the large $^{18}O/^{17}O $ given by Justtanont et al. (2012).
On the other hand, Justtannont et al. (2013) mentioned that there were problems with the observations and analysis of WX PsC, 
related to possible assymetries of the circumstellar shell, which may alter their result. In conclusion, any definite interpretation of
the evolutionary history of this star will be possible only when a more robust determination of the oxygen isotopic ratio will be available.

\subsection{Lithium abundances in O-rich AGB stars}
\citet{garcia07} presented results from high-resolution spectroscopy of a large sample
of O-rich AGB stars, for which the lithium and zirconium abundances were measured.
The latter element increases under the effects of TDU, thus
its content can be used as a reliable indicator of the efficiency of TDU in the stars
observed.

Among the sources observed by \citet{garcia07}, 25 show evidence of lithium, with
$\log \epsilon (^7Li) > 0.5$, whereas in 32 of them the lithium line was not
detected, thus indicating that $\log \epsilon (^7Li) < -1$. The distribution of the 
periods observed is $350-1200$ d for lithium-rich objects, whereas the AGB stars with
no lithium have periods below 500 d, with the single exception of
IRAS 18050-2213, which has a period of 732 d.

The lithium-rich stars in the \citet{garcia07} sample are interpreted as the progeny of
$M_{init} \geq 3.5~\Msun$ stars, currently evolving through the lithium-rich phase, 
when the Cameron-Fowler mechanism is active. Based on the discussion in section \ref{litio},
we know that this phase extends for about half of the AGB evolution of stars of solar
metallicity. While on general grounds we cannot identify the mass of the progenitors,
statistical arguments suggest that most of the lithium-rich stars descend from 
$4-5~\Msun$ stars. As shown in the right panel of Fig.~\ref{flitio}, the duration of the
lithium-rich phase is longer the lower is the mass of the progenitors: it is
$1.6\times 10^5$ yr for $M = 4~\Msun$, $8\times 10^4$ yr for $M = 5~\Msun$ and
$4\times 10^4$ yr for $M = 6~\Msun$. Given these time scales and the functional form 
of any realistic mass function, we deduce that the stars observed likely descend from
progenitors of mass below $5~\Msun$, and have current masses between $2~\Msun$ and  $5~\Msun$

The distribution of the periods of the stars in the sample by \citet{garcia07} further
supports this interpretation. The stars with no lithium are either stars of mass below
$3.5~\Msun$, which do not experience any HBB, or more massive objects in the initial
AGB phases, before the Cameron-Fowler mechanism is activated: during these early TP-AGB
phases the stars are more compact and less luminous, thus their periods are shorter.
This is fully consistent with one of the main results of the Garcia-Hernandez et al. (2007) analysis, i.e. lithium-rich stars have larger periods than their lithium-rich counterparts.
The lack of any strong s-process enrichment in the lithium-rich stars observed by
\citet{garcia07}, as deduced by the absence of significant zirconium enrichment, further supports
our models. Indeed this is in agreement with our results, that TDU is scarcely efficient 
in solar metallicity, massive AGB stars (see the values of $\lambda$ reported in Table 1).

%Because these stars show up lithium-rich chemistries for a significant fraction of the AGB  phase, the range of present masses potentially involved in this sample is not narrow, rather it extends from $\sim 3.5~\Msun$ to $\sim 8~\Msun$; however, for the arguments given above, we expect that the vast majority, if not all, of the lithium-rich stars in the \citet{garcia07} sample have mass $3.5~\Msun < M < 5~\Msun$.

A final comment concerns the luminosities of lithium-rich stars. Because the ignition 
of the Cameron-Fowler mechanism requires a minimum temperature at the bottom of the
envelope $T_{bce}\sim 30$ MK, this reflects into a minimum luminosity 
$L=1.8\times 10^4 L_{\odot}$, i.e. $M_{bol} = -5.88$; this stems from the tight
relationship between $T_{bce}$ and $L$. We note that, although Galactic massive
HBB-AGB stars may display strongly variable luminosities and their distances are
unknown \citep{garcia07}, similar truly massive HBB-AGB stars in
the Magellanic Clouds consistently display extremely high luminosities of $M_{bol} <
-6$ \citep{garcia09}.

\subsection{C isotopes in different types of AGB stars from radio transitions}
The group of Galactic AGB stars by \citet{sample2} are the most complete sample presented so
far, with $^{12}C/^{13}C$ ratios available for stars in different
phases of the AGB evolution. This sample includes both carbon stars and oxygen-rich
objects. The results are based on radiative transfer modelling of the observed $^{12}CO$ 
and $^{13}CO$ radio transitions; the solution of the
energy balance equation allowed the determination of the circumstellar $^{12}CO/^{13}CO$,
the rate of mass loss and the expansion velocity. These information can be used to 
constrain the evolutionary models. 

Before entering the discussion, we believe important to stress at this point that the
observational data in the optical/near-IR are more representative of the
photosphere, while the radio data, such as those presented in this section, trace
the chemistry of the circumstellar
envelope. This is confirmed by recent results, showing that in some cases
both values do not agree (e.g., Vlemmings et al. 2013). Furthermore,
the interpretation of the radio data is subject to several assumptions and modelling.  
Typically, it is assumed that the radio transitions are optically thin and the 
$^{12}CO/^{13}CO$ flux ratio is equivalent to the $^{12}C/^{13}C$ ratio, which is
not always the case; in case that the radio transitions are optically
thick, the real $^{12}C/^{13}C$ ratio is generally underestimated. 

We will discuss the stars in the sample separately,
according to their being M- or C-star. We do enter into the discussion of the possible 
origin of J stars, i.e. the carbon-rich objects in the sample with unusually low (below 15) 
$^{12}C/^{13}C$ ratios: these sources, as discussed in section 6.2.3, likely belong to binary systems, thus they cannot 
be understood on the basis of the single star models used here.

\subsubsection{O-rich M-type AGB stars}
The carbon ratio of these objects ($^{12}CO/^{13}CO \sim 6-7$) exhibits the 
signature of HBB, tracing the equilibria of proton capture nucleosynthesis. As shown 
in Fig.~\ref{fratco} (see left panel), this is a common behaviour of all the models of 
initial mass above $3~\Msun$ discussed here. 

The possibility that these objects descend from stars with mass just above 
the threshold required to activated HBB, i.e. $3.5~\Msun$, is unlikely, 
because these stars reach the surface $^{12}C/^{13}C$ corresponding to the equilibrium 
of proton capture nucleosynthesis only in the final TPs, thus for a limited fraction of
the AGB life (see the track corresponding to the $3.5~\Msun$ case in the
left panel of Fig.~\ref{fratco}). 

We believe more probable that IRC+10529 and IRC+50137 
descend from $M_{init} \geq 4~\Msun$ stars and are currently evolving through the AGB
phases following the ignition of HBB. This hypothesis is supported by the optical depth 
given by the authors, $\tau_{10}=3$, indicating a large degree of obscuration, thus the 
presence of great amounts of silicate
dust in the wind. The mass loss rates indicated by \citet{sample2} (in the range 
$10^{-5} \Msun/yr < \dot M < 3\times 10^{-5} \Msun/yr$) rule out very massive progenitors, 
which shifts our attention towards $M_{init}\sim 4~\Msun$ objects. This conclusion is
further supported by statistical arguments, based on the duration of the AGB phase
of stars of different mass, reported in Table 1 and in the bottom, right panel of
Fig.~\ref{fagb}.\\ 
If this interpretation proves correct, the luminosity expected 
is $L \sim 2\times 10^4 L_{\odot}$, significantly higher than those adopted by the authors 
($L \sim 10^4 L_{\odot}$, see Table 3).

R Leo exhibits a surface $^{12}C/^{13}C = 6$, similar to IRC+10529 and IRC+50137, 
indicating that the surface material was exposed to CN cycling. Unlike 
IRC+10529 and IRC+50137, the star is not heavily obscured ($\tau_{10}=0.03$) and the given 
mass loss rate ($10^{-7} \Msun /yr$) is a factor of $\sim 100$ smaller. \\
The possibility that R Leo is currently experiencing HBB
is not supported by the latter two evidencies, unless it is currently evolving through
a phase when mass loss and dust production are temporarily interrupted. If this is the 
case, the luminosity should be not below $L \sim 2\times 10^4 L_{\odot}$, almost a factor 
10 higher than the value indicated by \citet{sample2}. \\
An alternative possibility is that this star descends from a $M_{init} \sim 1.5-2~\Msun$ 
progenitor and is currently evolving through the initial AGB phases, before becoming a
carbon star. Cool bottom burning 
during RGB ascending might account for the reduction of $^{12}C/^{13}C$: this process, 
proposed by \citet{boothroyd99}, is originated by deep circulation mixing below the base 
of the convective envelope, and has the effects of mixing material enriched in $^{13}C$ 
and depleted in $^{12}C$ to the surface. A problem with this interpretation is 
that the observed $^{12}C/^{13}C$ is smaller than the lowest predictions from cool
bottom burning modelling, i.e. $^{12}C/^{13}C \sim 10$.

GX Mon, IK Tau, IRC-30398, IRC+10365 share several properties in common with IRC+10529 
and IRC+50137: the measured $^{12}C/^{13}C$ shows up the effects of HBB and the optical 
depths, in the
range $0.5 < \tau_{10} < 1$, trace the presence of significant quantities of silicate
dust in the circumstellar envelope. We discuss these 4 stars separately, because the
carbon ratios given by the authors, $^{12}C/^{13}C \sim 10$, are higher than expected
on the basis of a pure CNO equilibria, although the errors associated to individual
abundances are compatible with a pure HBB chemistry. In the latter case the interpretation
of these sources would be similar to what was proposed earlier in this section
for IRC+10529 and IRC+50137.\\
Alternatively, the large degrees of obscuration and carbon isotopic ratios 
$^{12}C/^{13}C \sim 10$ are obtained in the final AGB phases of $M_{init} \sim 4~\Msun$
stars (see the $4~\Msun$ track in the left panel of Fig.~\ref{fratco}): the 
interpretation of these stars would be similar to the scenario proposed for 
AFGL 5379 and OH26.5+0.6, in section \ref{just}. The effective temperatures of the
stars in the late AGB phases, $T_{eff} \sim 2200$ K, are only slightly in excess of 
the temperatures indicated by \citet{sample2}. \\
According to this scenario, the stars in this group should  have present masses of the
order of $\sim 1-1.5~\Msun$. The expected luminosity is $L \sim 2\times 10^4 L_{\odot}$, 
a factor of 2 higher than proposed by \citet{sample2}.

CIT4, IRC+60169 and IRC+70666 have $^{12}C/^{13}C$ ratios in the range 
$20 < ^{12}C/^{13}C < 60$. They exhibit a 
significant degree of obscuration, with $0.3 < \tau_{10} < 1$, revealing the presence of 
silicate dust in the wind. While the large $^{12}C/^{13}C$'s indicate the effects of TDU, 
the presence of significant quantities of dust in the circumstellar envelope suggests 
advanced AGB stages of stars with progenitors of mass above $\sim 3.5~\Msun$: indeed
lower mass stars reach the C-star stage, and little dust formation occurs in the early
AGB phases, when the star is still oxygen-rich.\\
The carbon ratios and the mass loss rates proposed by \citet{sample2} are reproduced
by models with initial mass $M_{init} \sim 3.5~\Msun$, just above the threshold to 
activate HBB; as shown in Fig.~\ref{fratco} (see left panel), these stars first experience
a series of TDU events, favouring the increase in the surface $^{12}C$, then produce
$^{13}C$ via HBB. If this understanding is correct, the stars should have a current
mass of $\sim 2.5~\Msun$ and a luminosity $L \sim 1.5\times 10^4 L_{\odot}$. \\
Alternatively, the degree of obscuration and the rate of mass loss proposed are reproduced 
by models of initial mass $4-4.5~\Msun$, in the final evolutionary phases: similarly to the
stars discussed in the previous point, the large $^{12}C/^{13}C$ might be the effect of late TDU episodes, occurring when HBB is turned off. If this interpretation is 
correct, we may fix the current mass and luminosity of these stars, that are, 
respectively, $M \sim 1.5~\Msun$ and $L \sim 2\times 10^4 L_{\odot}$.\\
In both cases the luminosities expected are significantly in excess of the suggestion 
by the authors, that give $L = 4000 L_{\odot}$ for IRC+60169 and IRC+70666.

SW Vir and RX Boo have $^{12}C/^{13}C \sim 20$, which is compatible with
the chemistry of any star at the beginning of the AGB phase, when the chemical
composition was modified solely by the first and, possibly, the second dredge-up
episodes. \\
Based on the luminosities given by the authors, $L = 4000 L_{\odot}$, we conclude that
SW Vir and RX Boo descend from $1.5 - 2.0~\Msun$ objects and are currently at the 
beginning of the AGB evolution, before reaching the C-star stage. This interpretation is
also in agreement with the very small optical depths given by \citet{sample2}, 
$\tau_{10} \sim 0.02-0.03$.\\ 
However, the same isotopic ratios and small degree of obscuration are also reproduced by
higher mass models in the early AGB phases, before the ignition of HBB. In this case
the luminosities would be $L \sim 10^4 L_{\odot}$, larger than the values given
by \citet{sample2}.

Concerning W Hya, R Dor, RT Vir, R Cas, the material in the surface regions of these 
stars were exposed to partial CN 
cycling, as confirmed by the observed isotopic carbon ratios, $^{12}C/^{13}C \sim 10$. 
The luminosities given by the authorsfor these stars are   in the range $4\times 10^3 L_{\odot} < L < 6\times 10^3 L_{\odot}$
%are based on accurate parallax  determination, provided by Hipparcos. These luminosities 
If these luminosities will be confirmed by precise distance measurements (see next section) the possibility that the observed $^{12}C/^{13}C$'s are determined by HBB would be ruled out beacuse significantly smaller than 
those reached by the stars experiencing HBB.  
A valid alternative is that the stars in this group  descend from low-mass progenitors: the main arguments  
supporting this conclusion are: a) all the stars of mass in the range 
$1~\Msun < M_{init} < 3~\Msun$ evolve at luminosities similar to those observed, in
the AGB phases previous to the increase in the surface $^{12}C$ via TDU (the latter
mechanism would increase the $^{12}C/^{13}C$, far above the observed values); 
b) the degree of obscuration and the mass loss rate indicated by \citet{sample2}
are very small, which is typical of the AGB evolution of low-mass stars, before the
C-rich phase is reached.\\ 
On the theoretical side, we expect that the surface 
$^{12}C/^{13}C$ of these objects is modified by the first dredge-up, after
which according to standard modelling of mixing we have $^{12}C/^{13}C \sim 20$; 
this is a factor $\sim 2$ higher than observed. A solution for this discrepancy could be 
that the stars in this group experienced cool bottom processing during the RGB ascending 
\citep{boothroyd99}.

\subsubsection{C-rich N-type AGB stars}
LP And, V Cyg, CW Leo, RW LMi, V384 Per, UU Aur have a surface C/O above unity, 
the signature of repeated TDU events.
The mass loss rate is correlated to the surface $^{12}C/^{13}C$, which spans the
range $40 < ^{12}C/^{13}C < 100$; this is what we
expect from the AGB evolution of low-mass stars, as discussed in section
\ref{lowmass}. These 6 objects are therefore experiencing advanced evolutionary
phases of the AGB life, after becoming carbon stars. 

The observed $^{12}C/^{13}C$'s are attained by all the star with initial mass in the range 
$1.5~\Msun \leq M_{init} \leq 3~\Msun$, although the luminosities and
the mass loss rates given by \citet{sample2} suggest $1.5-2~\Msun$ progenitors. 
The optical depths given by the authors are in the range $0.2 < \tau_{10} < 1$,
which indicates the presence of significant quantities of carbon dust in the wind;
this is expected on the basis of AGB+dust modelling of stars evolving through the
C-star phase \citep{flavia14}.

The interpretation of UU Aur and V Cyg poses some problems. UU Aur has the largest 
$^{12}C/^{13}C$ in the overall sample, namely $^{12}C/^{13}C=100$. Such large carbon 
abundances are reached by all the low-mass stars models considered here. The luminosity 
of this object is an issue
though: while according to our modelling the C-star stage is not reached as far as the 
luminosity is below $\sim 8000L_{\odot}$, \citet{sample2} indicate $L=4000L_{\odot}$.

The luminosity of V Cyg given by \citet{sample2}, $L=6000L_{\odot}$, is also not 
reproduced by our models. If confirmed, the low luminosities of these two objects
would be a strong indication that TDU is more efficient in the AGB stars of solar 
metallicity, compared to the predictions given here. 
%A reliable measurement of the distance of these object is urgently needed.

\subsubsection{S-type AGB stars}
The sample by \citet{sample2} includes 17 S-type stars, with a surface C/O around
unity. The interpretation of these objects is not straightforward, because the
$^{12}C/^{13}C$ and the luminosities given in the above paper are not consistent
with our predictions. Concerning the chemical composition, Fig.~2 in \citet{sample2}
shows that the average $^{12}C/^{13}C$ of this group of stars is slightly above 20,
whereas according to our models carbon stars should have $^{12}C/^{13}C > 50$. This
is shown in Fig.~\ref{fco} (right panel); note that the same chemistry is also expected
on the basis of C15 models, which adds more robustness to this general conclusion.
This systematic difference can be explained only by invoking some ad hoc mechanism,
such as cool bottom burning, acting to increase the $^{13}C$ in the envelope of low-mass 
stars before they become enriched in $^{12}C$; we believed this possibility unlikely 
though, because all the S-stars in the \citet{sample2} sample present such a low 
$^{12}C/^{13}C$, thus indicating that this mechanism should be active in all low-mass
stars. Alternatively, the circumstellar $^{12}C/^{13}C$ is not a reliable tracer of the
surface cratio, at least for S-type stars. The interpretation of the results by 
\citet{sample2} is further complicated by the differences among the luminosities
expected based on our models and those given by the authors. As discussed in section
\ref{lowmass}, and shown in Fig.~\ref{fco}, carbon stars of solar chemistry are expected 
to evolve at luminosities $L > 8000L_{\odot}$. This is at odds with the luminosities
adopted by \citet{sample2} (see their Table 1), which are in the range 
$4000L_{\odot}-12000L_{\odot}$. Note that use of C15 models (shown in the same figure)
would hardly improve this mismatch, as in that case luminosities not below $7000L_{\odot}$ 
are expected. 
\subsection{Distance estimates within  Gaia mission}
This detailed analysis shows that reliable measurements of the distance of 
Galactic AGB, especially of those with recent estimates of CNO elemental and 
isotopic abundances, is urgently needed. The knowledge of the distance will allow 
a robust determination of the luminosity, which, 
as discussed in the previous sections, is crucial to the characterization  of the 
observed stars in term of mass and evolution.\\
In Table 3 we summarized the characteristics of the 
Ramstedt \& Olofssson (2014) sample discussed in details in subsection 6.3. The authors
report the predicted absolute luminosity, however only in a few cases (starred with an 
asterisk in the table) the luminosities were estimated from accurate Hipparcos parallax
measurements or by VLBI maser spot astrometry. In all other cases, the luminosity was 
either derived from Groenewegen \& Whitelock (1996) period-luminosity relation (Mira 
variables) or assumed to be equal to 4000L$_{\odot}$ (semi-regular, irregular variables, 
variables of unknown type or period).
The uncertainty in the observed luminosity estimates makes the comparison between 
observed and predicted luminosities inconclusive.\\
As mentioned in the introduction, this problem will be addressed when Gaia astrometry for 
these stars will be available \footnote{Seven stars of the Ramstedt \& Olofssson (2014) sample 
will probably already be included in Gaia's first data release, foreseen by the end of 
summer 2016, which will include parallaxes for the large majority of the Hipparcos and Tycho-2 stars.}.
The accuracy of Gaia parallaxes depends in a complicated way on several factors: 
number of observations, environment (i.e. stellar density), brightness, colour and so on.
The number of end-of-mission observations based on Gaia scanning law\footnote{Computed with 
the Observation Forecast Tool available at$http://gaia.esac.esa.int/gost/index.jsp$}  
are reported in the last column of Table 3. 
A good fraction of the Ramstedt \& Olofssson (2014) sample stars will probably be observed 
enough times to reach the nominal 
astrometric error for bright stars of their spectral type 
($\sim$$\sigma_{\pi}$=10 $\mu$as). However, given that most of the stars in our sample 
are Mira or semi-regular variables, it is not possible at present to evaluate the actual 
parallax accuracy for them. If accurate parallaxes will be available, then it will be 
possible to derive accurate luminosities, to discriminate among different model scenarios 
and to assign an evolutionary mass in several cases.

\begin{table}
\begin{center}
\caption{Sample sources discussed in detail in subsection 6.3 with their spectral  
type (S-type), the derived $^{12}C/^{13}C$ (assumed to be equal to the measured 
$^{12}CO/^{13}CO$) and computed absolute magnitude given by 
Ramstedt \& Olofsson (2014). In the last column are shown  the  predicted  number of 
end-of-mission Gaia observations. Stars with  parallax measurements by Hipparcos   
or by more precise estimates as VLBI maser spot astrometry are labelled with an asterisk.} 
\begin{tabular}{l|l|l|l|l}
\hline
\hline 
Name& S-type  & $^{12}C/^{13}C$ & L/L$_{\odot}$ & N$_G$\\
\hline
IRC+10529   &M &7   &10600 & 57\\
IRC+50137   &M &6   & 9900& 46\\
R Leo*      &M &6   &2500 &30 \\
GX Mon      &M &11  &8200&21 \\
IK Tau      &M &10  &7700 & 48\\
IRC-30398   &M &13  &8900 & 21\\
IRC+10365   &M &13  &7700 & 39 \\
CIT4        &M &29  &4000 & 57\\
IRC+60169   &M &29  &4000 & 82\\
IRC+70066   &M &66  &4000 &76 \\
SW Vir*     &M &18   &4000 &45\\
RX Boo*     &M &17  &4000 &47\\
W Hya*      &M &10   &6000& 24\\
R Dor*      &M &10   &4000 & 36 \\
RT Vir*     &M & 9   &4500 & 39 \\
R Cas*      &M &19   &4000 & 82\\
LP And      &C &56   &9600 &68 \\
V Cyg       &C &38  &6000 &47 \\
CW Leo      &C &71  &9800 & 27 \\
RW Lmi      &C &45  &10000 & 52 \\
V384 Per    &C &43  &8300 & 43\\
UU Aur      &C &100  &4000 & 22\\
\hline 
\hline 
\end{tabular}
\end{center}
\label{tabrates}
\end{table}
\label{dustmodel}

\section{Conclusions}
We present solar metallicity models of the AGB phase of stars with mass in the
range $1~\Msun < M < 8~\Msun$. This investigation integrates previous explorations
by our group, focused on sub-solar chemistries.

The main physical and chemical properties of AGB stars are extremely sensitive to the 
stellar mass. A threshold mass $M \sim 3-3.5~\Msun$ separates two distinct behaviours.

The chemical composition of stars of mass $M \leq 3~\Msun$ is altered by the TDU
mechanism, which favours a gradual increase in the surface carbon content. We find that
the stars with mass in the range $1.5~\Msun \leq M \leq 3~\Msun$ become carbon stars
during the AGB phase. Once the C-star stage is reached, the consumption of the envelope
is accelerated by the expansion of the external regions and by the effects of radiation 
pressure acting on the carbonaceous dust particles in the circumstellar envelope. This
effects prevent further significant enrichment in the surface carbon, keeping the 
C/O ratio below $\sim 1.5$. The gas ejected by these stars is enriched in carbon
and nitrogen by a factor $\sim 3$ compared to the material from which the stars formed.
The luminosities of carbon stars fall in the range 
$8\times 10^3L_{\odot} < L < 1.2\times 10^{4}L_{\odot}$.

Stars of mass $M > 3~\Msun$ experience HBB at the bottom of the convective envelope.
The strength of the HBB increases with the mass of the star. The pollution 
from these stars reflects the equilibrium abundances of the HBB nucleosynthesis 
experienced. On general grounds, we expect carbon-poor and nitrogen-rich ejecta,
owing to CN cycling. In stars of mass above $\sim 5~\Msun$ the HBB temperatures are
sufficiently large to activate the full CNO and the Ne-Na nucleosynthesis: the gas
expelled by these stars is enriched in sodium, whereas the oxygen content is smaller
than it was when the star formed. These stars are expected to evolve as lithium-rich sources
for a significant fraction of the AGB phase.

The comparison with results in the literature outlines some similarities but also
significant differences, particularly for what regards the strength of the HBB
experienced, thus the luminosities at which these stars evolve and the kind of
pollution expected. The carbon, nitrogen and sodium content of stars of mass 
above $3~\Msun$ are extremely different from the results from other research teams, 
stressing the importance of confirmation from the observations.

We compare the models presented here with the CNO elemental and isotopic abundances in different types of Galactic AGB stars as estimated from observational data at very different wavelengths (from the optical to the radio domain);this part of the research has the double scope of 
adding more robustness to the present results and to characterise to stars observed, in
terms of mass and age of the progenitors. The comparison with the observations is
hampered by the unknown distances of the sources discussed.

\section*{Acknowledgments}
MDC acknowledges the contribution of the FP7 SPACE project ASTRODEEP (Ref.No:312725), 
supported by the European Commission. DAGH was funded by the Ram\'on y Cajal fellowship 
number RYC$-$2013$-$14182 and he acknowledges support provided by the Spanish Ministry of 
Economy and Competitiveness (MINECO) under grant AYA$-$2014$-$58082-P.
FD acknowledges support from the Observatory of Rome.

\end{document}